\documentclass[letterpaper,12pt]{JHEP3}
\pdfoutput=1
\usepackage{amsmath}
\usepackage{amssymb}
\usepackage{graphicx}
\usepackage{subfig}
\usepackage{bbm}
\usepackage{epsfig}
\usepackage{yfonts}
\newcommand{\labell}[1]{\label{#1}}
\def\({\left(} \def\){\right)}
\def\[{\left[} \def\]{\right]}
\def\al{\alpha} 
\def\del{{\partial}}

\def\hz{\mathrel{\mathop h^{\scriptscriptstyle{(0)}}}{}\!\!}

\def\gz{\mathrel{\mathop g^{\scriptscriptstyle{(0)}}}{}\!}
\def\go{\mathrel{\mathop g^{\scriptscriptstyle{(1)}}}{}\!}
\def\ho{\mathrel{\mathop h^{\scriptscriptstyle{(1)}}}{}\!}

\def\gs{\mathrel{\mathop g^{\scriptscriptstyle{(2)}}}{}\!\!\!}

\def\gn{\mathrel{\mathop g^{\scriptscriptstyle{(n)}}}{}\!\!\!}
\def\hn{\mathrel{\mathop h^{\scriptscriptstyle{(n)}}}{}\!\!\!}

\newcommand\I[1]{I^{\scriptscriptstyle{(#1)}}}

\newcommand{\non}{\nonumber \\}

\newcommand{\be}{\begin{equation}}
\newcommand{\ee}{\end{equation}}
\newcommand{\bea}{\begin{eqnarray}}
\newcommand{\eea}{\end{eqnarray}}
\newcommand{\ba}{\begin{eqnarray}}
\newcommand{\ea}{\end{eqnarray}}

\newcommand{\beq}{\begin{equation}}
\newcommand{\eeq}{\end{equation}}
\newcommand{\beqa}{\begin{eqnarray}}
\newcommand{\eeqa}{\end{eqnarray}}
\newcommand{\beqar}{\begin{eqnarray*}}
\newcommand{\eeqar}{\end{eqnarray*}}

\newcommand{\reef}[1]{(\ref{#1})}

\newcommand{\eg}{{\it e.g.,}\ }
\newcommand{\ie}{{\it i.e.,}\ }
\newcommand{\comment}[1]{{\bf [[[#1]]]}}

\newcommand{\mt}[1]{\textrm{\tiny #1}}

\newcommand{\veps}{\varepsilon}

\newcommand{\A}{\mathcal{A}}

\newcommand{\K}{\mathcal{K}}

\newcommand{\R}{\mathcal{R}}

\newcommand{\tL}{\tilde{L}}

\newcommand{\la}{\lambda}

\newcommand{\fin}{f_\infty}

\newcommand{\ct}{C_{T}} 
\newcommand{\hi}{{\hat \imath}}
\newcommand{\hj}{{\hat \jmath}}

\newcommand{\ads}{A} 

\newcommand{\m}{\sigma} 

\newcommand{\tg}{\tilde{g}}
\newcommand{\tl}{\tilde{\ell}}
\newcommand{\hh}{\tilde{h}}
\newcommand{\tS}{\tilde{\Sigma}}
\newcommand{\tR}{\tilde{R}}

\preprint{arXiv:1304.nnnn [hep-th]}

\title{On Spacetime Entanglement}

\author{Robert C. Myers,$^{1}$ Razieh Pourhasan$^{1,2}$ and
Michael Smolkin$^{1}$\\
\it
$^1$Perimeter Institute for Theoretical Physics,\\
\ Waterloo, Ontario N2L 2Y5, Canada\\
$^2$Department of Physics \& Astronomy and Guelph-Waterloo Physics Institute,\\
\ University of Waterloo, Waterloo, Ontario N2L 3G1, Canada}

\abstract{We examine the idea that in quantum gravity, the
entanglement entropy of a general region should be finite and the leading
contribution is given by the Bekenstein-Hawking area law. Using holographic
entanglement entropy calculations, we show that this idea is realized in the
Randall-Sundrum II braneworld for sufficiently large regions in smoothly curved
backgrounds. Extending the induced gravity action on the brane to include the
curvature-squared interactions, we show that the Wald entropy closely matches
the expression describing the entanglement entropy. The difference is that
for a general region, the latter includes terms involving the
extrinsic curvature of the entangling surface, which do not appear in the Wald
entropy. We also consider various limitations on the validity of these results.}

\begin{document}

\section{Introduction}

Considerations of the second law of thermodynamics in the presence of black
holes, led Bekenstein\cite{beks} to make the bold conjecture some forty years
ago that black holes carry an intrinsic entropy given by the surface area of
the horizon measured in Planck units multiplied by a dimensionless number of
order one. This conjecture was also supported by Hawking's area theorem
\cite{area}, which shows that, like entropy, the horizon area can never
decrease (in classical general relativity). Bekenstein offered arguments for
the proportionality of entropy and area, which relied on information theory, as
well as the properties of charged rotating black holes in general relativity
\cite{beks}. Of course, a crucial insight came with Hawking's discovery that
external observers around a black hole would detect the emission of thermal
radiation with a temperature proportional to its surface gravity \cite{hawk},
\ie $T=\frac{\kappa}{2\pi}$. Combining this result with the four laws of black
hole mechanics \cite{four}, the black hole entropy was recognized to be
precisely
\be S_{BH} = \frac{\cal A}{4G} \,, \labell{prop0} \ee
where $\cal A$ is the area of the event horizon. In fact, this expression
applies equally well to any Killing horizon, including de Sitter \cite{DS} and
Rindler \cite{ray} horizons. While originally derived with considerations of
general relativity in four spacetime dimensions, eq.~\reef{prop0} also
describes the entropy for black hole solutions of Einstein's equations in
higher dimensions.\footnote{In $d$ spacetime dimensions, the `area' has units
of $\text{\emph{length}}^{d-2}$.} Further, it has been shown that the
Bekenstein-Hawking (BH) expression \reef{prop0} can be extended to a general
geometric formula, the `Wald entropy', to describe the horizon entropy in
gravitational theories with higher curvature interactions \cite{WaldEnt}.

Of course, much of the interest in black hole entropy, and black hole
thermodynamics, stems in the hope that it provides a window into the nature of
quantum gravity. A recent conjecture \cite{new1} proposes the above area law
\reef{prop0} has much wider applicability and serves as a characteristic
signature for the emergence of a semiclassical metric in a theory of quantum
gravity.\footnote{See also discussion in \cite{luty}.} The precise conjecture
was that in a theory of quantum gravity, for any sufficiently large region in a
smooth background spacetime, the entanglement entropy between the degrees of
freedom describing a given region with those describing its complement is
finite and to leading order, takes the form given in eq.~\reef{prop0}. Of
course, an implicit assumption here is that the usual Einstein-Hilbert action
(as well as, possibly, a cosmological constant term) emerges as the leading
contribution to the low energy effective gravitational action. This conjecture
was supported by various lines of evidence: First of all, in the context of
gauge/gravity duality, eq.~\reef{prop0} is applied to general surfaces in
evaluating holographic entanglement entropy \cite{rt1}. Second, it can be shown
that in perturbative quantum field theory, the leading area law divergence
\cite{arealaw} appearing in calculations of the entanglement entropy for a
general region $V$ can be absorbed by the renormalization of Newton's constant
in the BH formula applied to the boundary of $V$, \ie with the area
$\A(\partial V)$. These arguments are framed in terms of the entanglement
Hamiltonian describing the reduced density matrix and require understanding
certain general properties of the latter operator. However, this new
understanding can also be combined with Jacobson's `thermodynamic' arguments
\cite{magic} for the origin of gravity to provide further independent support
of the above conjecture. A preliminary calculation in loop quantum gravity also
provides support for this new idea. Finally, in models of induced gravity
\cite{sak}, certain results \cite{rob06,fur06} were again in agreement with the
idea that eq.~\reef{prop0} describes the entanglement entropy of general
regions, in particular even when the entangling surface does not coincide with
an event horizon.

In this paper, we study this conjecture in more detail in the context of
induced gravity. In particular, following \cite{rob06}, we will study
entanglement entropy in the Randall-Sundrum II (RS2) braneworld \cite{RS2} and
our main result is as follows: The induced gravity action on the brane takes
the form
 \be
I_{ind}=\int d^{d}x\sqrt{-\tg} \[ {R\over 16\pi  G_{d}}
+\,\frac{\kappa_1}{2\pi}\bigg(R_{ij}R^{ij}-{d\over 4(d-1)}R^2\bigg)
+\frac{\kappa_2}{2\pi} \,C_{ijkl} C^{ijkl}+\cdots  \] ~.
  \labell{EGBind00}
 \ee
where the various curvatures are calculated for the brane metric $\tg_{ij}$ and
the ellipsis indicates cubic and higher curvature interactions. The precise
value of the $d$-dimensional Newton's constant and the induced couplings of the
curvature-squared terms depend on the details of the dual bulk theory and we
determine these for two different examples. In principle, these calculations
can be extended to higher orders in the derivative expansion but as indicated
above, we ignore any contributions beyond curvature-squared. Then with
holographic calculations of entanglement entropy, we find for any sufficiently
large region $V$ on the brane, the corresponding entanglement entropy is given
by
 \bea
S_\mt{EE}&=&\frac {{\cal A}(\tS)}{4 G_{d}} +\kappa_1\int_{\tS} d^{d-2}y
\sqrt{  \hh} \[ 2R^{ij}\, {\tg}^{\perp}_{ij}-{d\over d-1}\,R-K^iK_i
\]
\labell{entJM0}\\
&&\quad +\ 4\kappa_2\int_{\tS}d^{d-2}y \sqrt{  \hh}
\[\hh^{ac}\hh^{bd}C_{abcd}-K^i_{ab}K_i{}^{ab}+\frac{1}{d-2}K^iK_i\]
+\cdots \, , \nonumber
 \eea
where $\hh_{ab}$ and $K^i_{ab}$ are, respectively, the induced metric and the
second fundamental form of the entangling surface $\tS=\partial V$. The leading
contribution here is captured by the Bekenstein-Hawking formula \reef{prop0},
in precise agreement with the conjecture of \cite{new1}. We can also compare
the above result with the Wald entropy \cite{WaldEnt} for the induced
gravitational action \reef{EGBind00}. Then we find that $S_\mt{EE}$ and
$S_\mt{Wald}$ also agree at this order in the derivative expansion, except that
the extrinsic curvature terms in eq.~\reef{entJM0} do not appear in the Wald
entropy. It is noteworthy that the coefficients of these additional terms are
still determined by the higher curvature couplings in the effective gravity
action \reef{EGBind00}. We should emphasize that our calculations only capture
the leading terms in an expansion for large central charge of the braneworld
conformal field theory.

An overview of the remainder of the paper is as follows: We begin a brief
review of the RS2 model as a theory of induced gravity, in section
\ref{ransun}. In section \ref{region}, we use holographic entanglement entropy
to evaluate $S_\mt{EE}$ for general regions on the RS2 brane, with the result
given in eq.~\reef{entJM0}. In section \ref{test}, we consider our results in
the context of various inequalities that the entanglement entropy must satisfy.
This comparison points out certain limitations with the present approach. Then
we conclude with a discussion of our results in section \ref{discuss}. A number
of appendices are included which describe various technical details. In
appendix \ref{effact}, we derive the induced gravity action on the brane for
the case when the dual bulk theory is described by Gauss-Bonnet gravity. Of
course, setting the curvature-squared coupling to zero in the previous result
yields the induced action for Einstein gravity in the bulk. Appendix \ref{geom}
considers in detail the geometry of the codimension-two surfaces in the bulk
and derives various expressions for the curvatures that are useful in deriving
the holographic entanglement entropy in section \ref{region}. In appendix
\ref{sphere}, we compare the perturbative results for the entanglement entropy
given in section \ref{region} with those for the simple case of a spherical
entangling surface in flat space where the entire holographic result can be
calculated analytically.

\section{Randall-Sundrum II} \labell{ransun}

In their seminal work \cite{RS2}, Randall and Sundrum  showed that standard
four-dimensional gravity will arise at long distances on a brane embedded in a
noncompact but warped five-dimensional background. Their construction starts by
taking two copies of five-dimensional anti-de Sitter (AdS) space and gluing
them together along a cut-off surface at some large radius with the three-brane
inserted at this junction. This construction readily extends to an arbitrary
number of spacetime dimensions to produce gravity on a $d$-dimensional brane
\cite{highRS} and in fact, it is straightforward to see that the braneworld
metric is governed by the full nonlinear Einstein equations in $d$ dimensions,
to leading order in a derivative expansion \cite{highRS}. Our understanding of
these Randall-Sundrum II (RS2) models is greatly extended by realizing the
close connection with the AdS/CFT correspondence --- see \cite{herman,gubser}
and references therein. Given the holographic description of AdS space, we have
a dual description of the braneworld which is entirely in $d$ dimensions,
namely, gravity, as well as any brane matter, coupled to (two copies of) a
strongly coupled CFT with a UV cut-off. Interestingly, in this context, we can
think of the RS2 model as a theory of induced gravity \cite{andy,rob06}.

Of course, the key difference between the standard AdS/CFT correspondence and
the RS2 model is that the bulk geometry is cut off at some finite
$\rho=\rho_c$, which gives rise to a new normalizable zero-mode in the bulk
gravity theory. This extra mode is localized at the brane position and becomes
the propagating graviton of the $d$-dimensional gravity theory. One may make
use of the calculations and techniques for regulating the bulk theory in
AdS/CFT correspondence \cite{counter,construct} to determine the action of the
induced gravity theory on the brane. We sketch this approach here and relegate
a detailed calculation of the boundary action to appendix
\ref{effact}.\footnote{Although the context is somewhat different, our approach
is similar in spirit to the discussion of boundary actions in \cite{aninda}.}
As a theory of $(d+1)$-dimensional gravity, the RS2 model has the following
action
 \be
 I_{RS}=2\,I_{bulk}+I_{brane}\,,
 \labell{act1}
 \ee
where $I_{bulk}$ is the bulk gravitational action\footnote{We introduced a
factor of two here as a reminder that there are two copies of the AdS
geometry.} and $I_{brane}$ includes contributions of matter fields localized on
the brane, as well as the brane tension. To determine the effective action for
the $d$-dimensional gravity theory on the brane, one needs to integrate out the
extra radial geometry in the AdS bulk. In the context of AdS/CFT
correspondence, one must introduce a cut-off radius\footnote{We will assume
that $\rho=0$ corresponds to the AdS boundary
--- see section \ref{region} from more details.} $\rho=\rho_c$ to regulate
this calculation. Of course, in the RS2 model, this cut-off acquires a physical
meaning as the position of the brane and so the integral is naturally
regulated. The general result takes the form:
 \be
I_{bulk}=I_{fin}+\sum_{n=0}^{\lfloor d/2\rfloor}\I n \, ,
 \labell{bulkact}
 \ee
where each of the terms in the sum, $\I n$,
diverges as $\rho_c^{n-d/2}$ in the limit $\rho_c\to0$,\footnote{For even $d$,
the divergence is logarithmic for $n=d/2$.} while $I_{fin}$ is a non-local
contribution which remains finite in this limit. In fact, each $\I n$ is given
by an integral over the brane of a (local) geometric term constructed from the
boundary metric, its curvature and derivatives of the curvature. The label $n$
designates the number of derivatives appearing in the geometric term, \ie $\I
n$ contains $2n$ derivatives of the metric.

In the context of AdS/CFT correspondence, these expressions can be seen as
local divergences that result from integrating out the CFT degrees of freedom
with the regulator $\rho=\rho_c$. Boundary counterterms are added to precisely
cancel the $\I n$, allowing one to take the limit $\rho_c\to0$ with a finite
result for the gravitational action \cite{counter}. In the context of the RS2
model, the cut-off is fixed, no additional counter-terms are added and the
total action \reef{act1} becomes
 \be
 I_{ind}=2\sum_{n=0}^{\lfloor d/2\rfloor}\I n + 2I_{fin} + I_{brane} \, .
 \labell{bulkact2}
 \ee
Hence, the effective gravitational action on the brane is given by the sum of
the geometric terms $\I n$, which can be interpreted in terms of a standard
derivative expansion, \eg the $n=0$, 1 and 2 terms will correspond to the
cosmological constant term, the Einstein term and a curvature-squared term,
respectively. In Appendix \ref{effact}, we explicitly illustrate these ideas by
deriving these three terms for both Einstein and Gauss-Bonnet gravity in the
bulk. In this regard, the brane tension in $I_{brane}$ may be said to play the
role of a counter-term, in that we will tune the tension to precisely cancel
the $\I 0$ contribution so that the effective cosmological constant vanishes.
Further let us note that we must be working in a regime where the brane
geometry is weakly curved in order for the above derivative expansion to be
effective and for the local gravitational terms to dominate the $I_{fin}$
contribution --- see further details in section \ref{region}.

Above, the bulk cut-off $\rho=\rho_c$ plays an essential role in both the
AdS/CFT calculations and the RS2 model. Holography indicates that there is a
corresponding short-distance cut-off $\delta$ in the dual CFT. Again in the
AdS/CFT context, this is simply a convenient regulator and one imagines taking
the limit $\delta\to0$ after the appropriate counterterms are added. In the RS2
model, the cut-off remains fixed and one finds that $\delta=\tL$, \ie the
short-distance cut-off matches the AdS curvature scale in the
bulk.\footnote{Note that this result is independent of the choice of $\rho_c$.
Rather in the RS2 model, $\delta$ is defined in terms of the induced metric on
the brane. This should be contrasted with the standard AdS/CFT approach where
the CFT metric defining $\delta$ is the boundary metric rescaled by a factor of
$\rho_c$.} Therefore if $\delta$ is to be a small scale, then the bulk AdS
geometry is highly curved.

In fact, we can think of the RS2 model as having a single independent scale,
\ie the cut-off $\delta$. To illustrate this point, we focus on the case of
Einstein gravity in the bulk forthe following discussion.\footnote{As we will
see later, the situation for Gauss-Bonnet gravity is slightly more complicated.
In particular, the boundary CFT is characterized by two independent central
charges, both of which will be assumed to be large --- see eqs.~\reef{effectc}
and \reef{effecta}.} First of all, we saw that $\tL$ is fixed by $\delta$
above. Another scale in the bulk gravitational theory would be the Planck
scale, \ie $\ell^{d-1}_{P,bulk}\equiv 8\pi G_{d+1}$. The standard AdS/CFT
dictionary relates the ratio of the AdS curvature scale to Planck scale in
terms of a central charge $C_T$, which measures the number of degrees of
freedom in the boundary CFT. Hence in the RS2 model with $\delta=\tL$, we
define
 \be
\ct \equiv \pi^2\,\delta^{d-1}\!/\ell^{d-1}_{P,bulk}\,. \labell{centralT}
 \ee
Now the construction described above determines the induced couplings of the
brane gravity action \reef{bulkact2} in terms of the bulk Newton's constant (or
equivalently $\ell_{P,bulk}$) and the short-distance cut-off. Hence these
couplings can also be expressed in terms of $\delta$ and $C_T$. For example,
the effective Newton's constant \cite{highRS} (see also Appendix \ref{effact})
is given by
 \be
 G_d={d-2\over 2\,\delta}\, G_{d+1}= {\pi(d-2)\over 16}\,\frac{\delta^{d-2}}{\ct}\,.
 \labell{newton}
 \ee
Hence, in the RS2 model, both the bulk and boundary Planck scales are derived
quantities given in terms of $\delta$ and $C_T$, which we can regard as the
fundamental parameters defining the RS2 theory.

We must emphasize that throughout the following, we will assume that $C_T\gg1$
and our calculations only capture the leading terms in an expansion with large
$C_T$. First of all, this assumption is implicit in the fact that we will treat
the bulk gravity theory classically. Quantum corrections in the bulk will be
suppressed by inverse powers of $C_T$. Further, one must imagine that the
simple description of the RS2 model, with a discrete cut-off in the AdS bulk,
is an approximation to some construction within a UV complete theory, \eg a
stringy construction as described in \cite{herman,string}. In such a scenario,
the bulk cut-off will have a more elaborate realization, \eg where the AdS
space would extend smoothly into some compact UV geometry. Hence one should
expect that there will be additional contributions to the effective
gravitational action \reef{bulkact2}. Effectively, these can be catalogued as
additional counterterms (beyond the cosmological constant term) in $I_{brane}$.
However, it is reasonable to expect that these corrections should be
independent of the central charge defining the AdS contributions and so they
are again suppressed in the limit of large $C_T$. We might note that in the
limit $C_T\gg1$, we have $\delta\gg \ell_P$ for both the Planck scale in the
bulk and on the brane.

Finally, we observed above that the local terms in eq.~\reef{bulkact} can be
seen as being generated by integrating out the CFT degrees of freedom in the
context of the AdS/CFT correspondence. The same interpretation applies to the
RS2 model and so in this sense, this model \cite{rob06,andy} provides a theory
of induced gravity \cite{sak}. Such models received particular attention in
discussions of the idea that black hole entropy coincides with the entanglement
entropy between degrees of freedom inside and outside of the event horizon
\cite{induce}. In fact, \cite{rob06} used the RS2 model to illustrate this
idea. The approach taken there was to use the usual holographic prescription to
calculate entanglement entropy \cite{rt1}. That is, to calculate the
entanglement entropy between a spatial region $V$ and its complement $\bar V$
in the $d$-dimensional boundary theory, one extremizes the following expression
 \be
S(V) =\ \mathrel{\mathop {\rm ext}_{\scriptscriptstyle{\m\sim A}} {}\!\!}
\frac{{\A}(\m)}{4G_{d+1}}
 \labell{define}
 \ee
over ($d$--1)-dimensional surfaces $\m$ in the bulk spacetime, which are
homologous to the boundary region $V$.\footnote{Hence the `area' $\A(\m)$ to
denotes the ($d$--1)-dimensional volume of $\m$.} In particular then, the
boundary of $\m$ matches the `entangling surface' $\Sigma=\partial V$ in the
boundary geometry. While a general derivation of eq.~\reef{define} remains
lacking, there is a good amount of evidence supporting this proposal in the
context of the AdS/CFT correspondence, \eg see \cite{rt1,gbee,casini9,head}. In
\cite{rob06} and in the following, it is assumed that the same prescription
could be applied to the RS2 model. In an expansion for large $C_T$, it seems
reasonable to assume that $S(V)$ is dominated by correlations of the CFT
degrees of freedom and eq.~\reef{define} yields the leading contribution to the
entanglement entropy. In section \ref{test}, we discuss further limitations in
applying eq.~\reef{define} in the RS2 model.

The essential argument in \cite{rob06} was that in the RS2 model, extending the
event horizon of a black hole on the brane into the bulk naturally defines an
extremal surface in the AdS geometry. Hence if the entangling surface $\tS$ on
the brane coincides with the event horizon, eq.~\reef{define} simply evaluates
the expected black hole entropy. Similar, considerations were made for de
Sitter horizons for the RS2 braneworld in \cite{andy}. In \cite{rob06},
calculations were presented for an explicit black hole solution in a $d=3$
braneworld \cite{highRS} and it was shown that the leading contribution takes
the expected BH form \reef{prop0} for large black holes. However, it was also
noted that eq.~\reef{define} yields a finite entanglement entropy for a
circular entangling surface in empty (three-dimensional) Minkowski space and
further, the leading contribution is again ${\cal A}(\tS)/4G_{3}$, as long as
its radius satisfies $R\gg\delta$. In fact, it is straightforward to see that
the holographic prescription \reef{define} will yield a finite entanglement
entropy in any number of spacetime dimensions and for general entangling
surfaces in the RS2 model. We confirm, in the next section, that the leading
contribution takes precisely the form ${\cal A}(\tS)/4G_{d}$ for sufficiently
large regions, in agreement with the conjecture of \cite{new1}. Further, we
will examine the first higher curvature corrections to the BH expression
\reef{prop0}.


\section{Entanglement entropy for general regions} \labell{region}

In this section, we use the holographic prescription \reef{define} \cite{rt1}
and its generalization to Gauss-Bonnet gravity \cite{gbee,jan} --- see
eq.~\reef{eegb} --- to evaluate the entanglement entropy associated with
general entangling surfaces on the $d$-dimensional brane of the RS2 model. Our
calculations will make use of the Fefferman-Graham (FG) expansion \cite{feffer}
as developed to describe the boundary theory in the AdS/CFT correspondence
\cite{construct}. To begin, we write the asymptotic geometry of AdS space in
$d+1$ dimensions as\footnote{Let us comment on our index conventions throughout
the paper. Directions in the full (AdS) geometry are labeled with letters from
the second half of the Greek alphabet, \ie $\mu,\nu,\rho,\cdots$. Letters from
the `second' half of the Latin alphabet, \ie $i,j,k,\cdots$, correspond to
directions in the background geometry on the brane or on the boundary of AdS.
Meanwhile, directions along the entangling surface on the brane are denoted
with letters from the beginning of the Latin alphabet, \ie $a,b,c,\cdots$, and
directions along the corresponding bulk surface are denoted with letters from
the beginning of the Greek alphabet, \ie $\alpha,\beta,\gamma,\cdots$. Finally,
we use hatted letters from the later part of the Latin alphabet to denote frame
indices in the transverse space to both of these surfaces, \ie $\hi,\hj$.}
 \be
ds^2 = G_{\mu\nu}dx^\mu dx^\nu=\frac{\delta^2}{4}\frac{d\rho^2}{\rho^2} +
\frac{1}{\rho}\,g_{i j}(x,\rho)\,dx^i dx^j\,,
 \labell{expandfg}
 \ee
where $\delta=\tilde L$ is the AdS curvature scale and $\rho=0$ is the boundary
of AdS. Now the metric $g_{i j}(x,\rho)$ admits a series expansion in the
(dimensionless) radial coordinate $\rho$
 \bea
g_{i j}(x,\rho)&=& \gz_{i j}(x^i) + \rho \go_{i j}(x^i) + \rho^2 \gs_{i
j}(x^i)  + \cdots \,.
 \labell{expand}
 \eea
The leading term $\overset{\scriptscriptstyle{(0)}}{g}\!_{ij}$ corresponds to
the metric on the boundary of AdS space. The next set of contributions in this
expansion, \ie with $n<d/2$ (for either odd or even $d$), are covariant tensors
constructed from this boundary metric \cite{construct}. At higher orders $n\ge
d/2$, the coefficients $\overset{\scriptscriptstyle{(n)}}{g}\!_{ij}$ will also
depend on the specific state of the boundary CFT that is being described, \eg
$\langle T_{ij} \rangle$. However, in the context of AdS/CFT correspondence, it
was shown \cite{relevant} that only the coefficients with $n<d/2$ contribute to
the divergences appearing in the entanglement entropy of the dual CFT. As we
will see below, in the RS2 model, the analogous terms become the leading
contributions to the entanglement entropy. Moreover, rather than being
divergent, they can be expressed in terms of the couplings appearing in the
induced gravity action \reef{bulkact2}. These terms will be the focus of our
present calculations and so our results will be independent of the state of the
CFT.

In fact, the metric coefficients in the range $1\le n<d/2$ are almost
completely fixed by conformal symmetries at the boundary \cite{adam}. For
example, the first coefficient in the FG expansion in eq.~\reef{expand} is
independent of the details of the bulk gravity action and is given by
 \be
\go_{i j} = -\frac{{\delta}^2}{d-2}\bigg(R_{i j}[\gz]-
\frac{\overset{\scriptscriptstyle{(0)}}{g}_{ij}}{2(d-1)}R[\gz] \bigg)\,,
 \labell{metricexpand}
 \ee
where $R_{ij}$ is the Ricci tensor constructed with the boundary metric
$\overset{\scriptscriptstyle{(0)}}{g}\!_{ij}$. At higher orders, certain
constants (corresponding to coefficients of conformally covariant tensors) must
be fixed by the bulk equations of motion and so depend on the specific bulk
gravity theory. For example, for arbitrary
$\overset{\scriptscriptstyle{(0)}}{g}_{ij}$, the coefficient
$\overset{\scriptscriptstyle{(2)}}{g}_{ij}$ is given by \cite{adam}
 \bea
\gs_{i j} &=& \delta^4\,\bigg(k_1\,  C_{m n k l}C^{m n k l}\gz_{i j} +
k_2\, C_{i k l m}C_{j}^{~~k l m} \nonumber \\
&& + \frac{1}{d-4}\bigg[\frac{1}{8(d-1)}\nabla_i\nabla_jR - \frac{1}{4(d-2)}\Box R_{i j}+
\frac{1}{8(d-1)(d-2)}\Box R \gz_{i j} \nonumber\\
&&-\frac{1}{2(d-2)}R^{k l}R_{i k j l} + \frac{d-4}{2(d-2)^2}R_{i}^{~k}R_{j k}+
\frac{1}{(d-1)(d-2)^2}RR_{i j}\nonumber \\
&&+\frac{1}{4(d-2)^2}R^{k l}R_{k l}\gz_{i j}
-\frac{3d}{16(d-1)^2(d-2)^2}R^2 \gz_{i j}\bigg]\bigg)\,,
 \labell{metricexpand2}
 \eea
where $C_{mnkl}$ is Weyl tensor for the boundary metric. Above the two
constants, $k_1$ and $k_2$, will depend on the bulk gravity theory. For
example, they vanish with Einstein gravity in the bulk, while with Gauss-Bonnet
gravity they are given by eq.~\reef{k-coeff}.

In the RS2 model, the standard choice which we adopt is to set the position of
the brane at $\rho=\rho_c=1$. A scaling symmetry of the AdS geometry allows us
to make this choice without loss of generality. However, note that generally,
one thinks of the FG expansion, described by eqs.~\reef{expandfg} and
\reef{expand}, as being justified because it is applied in the vicinity of the
AdS boundary, \ie for $\rho\ll1$. Hence, some extra attention is required to
justify the FG expansion when it is applied in the RS2 model with the brane at
$\rho=1$. By a simple scaling argument,
$\overset{\scriptscriptstyle{(n)}}{g}\!_{ij}$ contains $2n$ derivatives with
respect to the boundary coordinates, as can be seen explicitly in
eqs.~\reef{metricexpand} and \reef{metricexpand2}. Hence we can regard the
expansion \reef{expand} as a derivative expansion and it will converge
effectively as long as the boundary metric
$\overset{\scriptscriptstyle{(0)}}{g}\!_{ij}$ is weakly curved on the scale of
the AdS curvature $\tL$, which in the RS2 models matches the short-distance
cut-off $\delta$ in the dual CFT. That is, we will require
 \be
\delta^2\,R^{ij}{}_{kl}[\gz]\ll 1\,, \labell{cond1}
 \ee
and similarly for (covariant) derivatives of the curvatures.\footnote{One
should imagine that the curvature is expressed in an orthonormal frame in this
inequality.} Further, we must keep in mind that the boundary metric
$\overset{\scriptscriptstyle{(0)}}{g}\!_{ij}$, which as we described above
determines the leading coefficients in the FG expansion \reef{expand}, does not
match the brane metric. Rather using eqs.~\reef{expandfg} and \reef{expand},
the induced metric on the brane is given by
 \be
\tg_{ij}= \left. G_{ij}\right|_{\rho=1} =g_{ij}(x,\rho=1) =
\gz_{ij}(x)+\go_{ij}(x)+
\cdots=\sum_{n=0}^\infty \gn_{ij}(x)~.
 \labell{bond}
 \ee
However, note that given the constraint \reef{cond1} on the boundary geometry
(and using eq.~\reef{metricexpand}), the differences between these two metrics
must be small since
 \be
\tg_{\,ij}\,-\gz_{ij}\sim \go_{ij}\ll1\,.
 \labell{okay}
 \ee

There is a similar (small) shift in the geometry of the entangling surface.
Standard calculations, \eg \cite{gbee,relevant,adam}, define the entangling
surface $\Sigma$ on the AdS boundary at $\rho=0$ --- see figure \ref{fgplot}.
Following the holographic prescription \reef{define}, one determines the
corresponding extremal surface $\sigma$ in the bulk. Now the entangling surface
$\tilde\Sigma$ on the brane is defined as the intersection of $\sigma$ with the
cut-off surface at $\rho=1$. Hence the geometries of these two surfaces will
not coincide but differences can be precisely determined using the FG
expansion, as we show in the following.
\FIGURE[!ht]{
\includegraphics[width=0.85\textwidth]{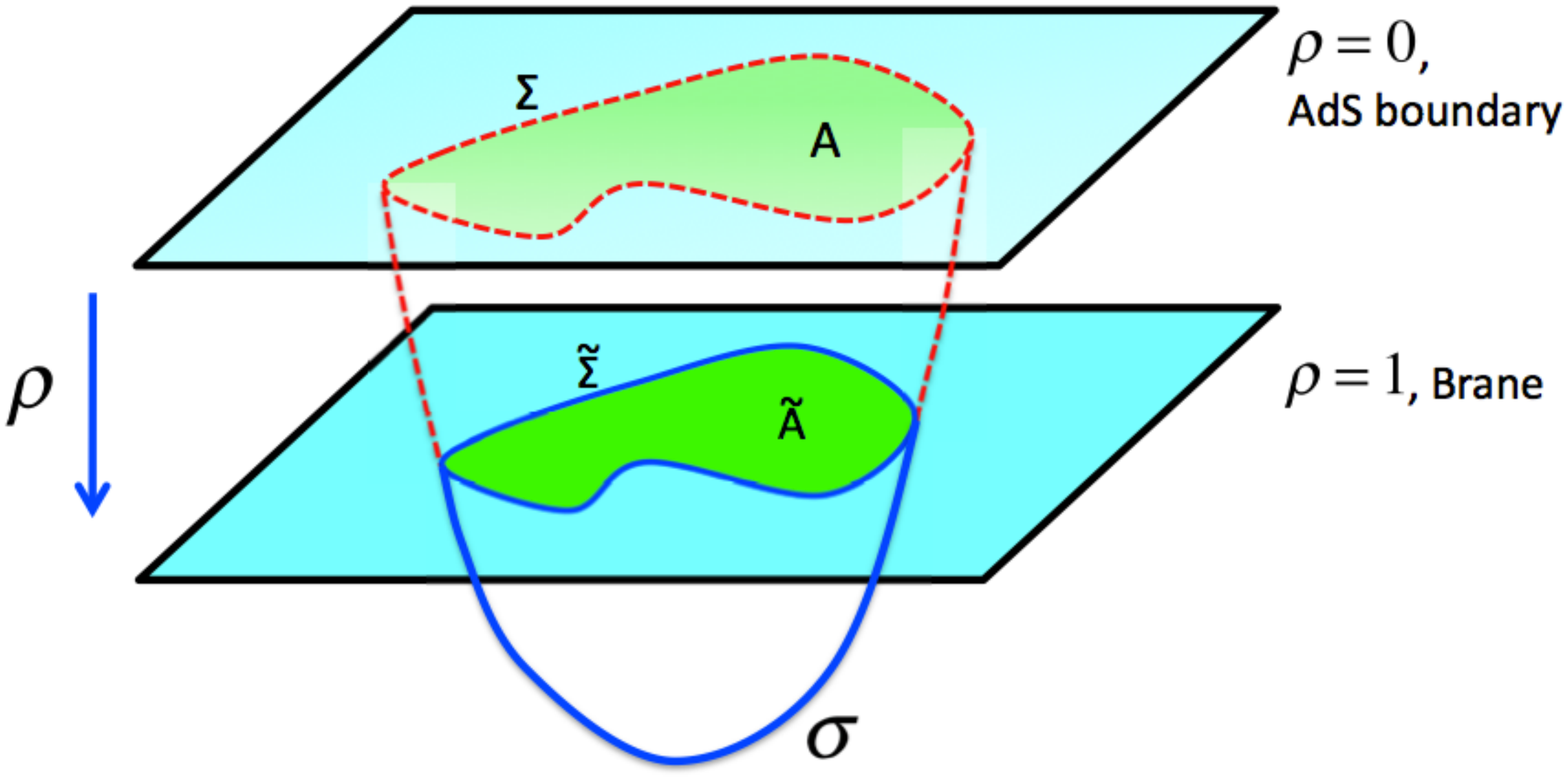}
\caption{(Colour Online) The entangling surfaces, $\Sigma$ on the AdS boundary
and $\tilde\Sigma$ on the brane, do not quite coincide because of the
nontrivial radial profile of the extremal surface $\m$ in the bulk.}
\labell{fgplot}}

Given the framework described above and shown in figure \ref{fgplot}, let $y^a$
with $a=1,\cdots,d-2$ be coordinates running along the entangling surface
$\Sigma$ in the AdS boundary and let $h_{\al\beta}$ be the induced metric on
extremal $\m$. Reparametrizations on this bulk surface are fixed by imposing
$h_{a\rho}=0$. In the same way that the FG expansion makes a Taylor series
expansion of the bulk metric in the vicinity of the AdS boundary, we can
represent the induced metric $h_{\al\beta}$ with a Taylor series about
$\rho=0$:\footnote{For further details, see appendix \ref{geom}.}
 \be
 h_{\rho\rho}={\delta^2\over4\rho^2}\Big(1+\ho_{\rho\rho}\,\rho+\cdots\Big)\,,\qquad
 h_{ab}={1\over \rho}\(\hz_{ab}+\ho_{ab}\,\rho+\cdots\)\, ,
\labell{inducemet}
 \ee
where $\overset{\scriptscriptstyle{(0)}}{h}_{ab}$ is the induced metric on the
entangling surface $\Sigma$. The first order coefficients in this expansion
again independent of the specific form of the bulk gravity action and are given
by \cite{gbee,relevant,adam}
 \be
\ho_{ab}=\go_{ab} -{ \delta^2\over d-2} K^i\,K^j_{ab}\gz_{ij}\,,\quad
\ho_{\rho\rho}={ \delta^2\over (d-2)^2}\,K^i\,K^j\gz_{ij}
 \,,
 \labell{indmetric}
 \ee
with $K^i_{ab}$ being the second fundamental form of $\Sigma$ (and $K^i=
\overset{\scriptscriptstyle{(0)}}{h}{}^{ab} K^i_{ab}$).\footnote{Here we adopt
the notation of \cite{adam}. Let $n^\hi{}_j$ (with $\hi=0,1$) be a pair of
orthonormal vectors which span the transverse space to $\Sigma$. The extrinsic
curvatures are then defined by $K^{\hi}_{ab}=\nabla_a n^\hi{}_b$ and
contracting with a normal vector gives $K^i_{ab}=n_\hj{}^{i }\,K^{\hj}_{ab}$.
Hence in the following formulae, the extrinsic curvatures carry a coordinate
index $i$, rather than a frame index $\hi$.}.

As above, we require that this expansion \reef{inducemet} is applicable in the
vicinity of the brane at $\rho=1$. The latter requires both that the background
curvatures are small as in eq.~\reef{cond1} but the characteristic scale of the
extrinsic curvatures is also much less than $\delta$, \ie
 \be
\delta\, K^i_{ab}\ll 1\, .
 \labell{cond2}
 \ee
Analogous inequalities would also have to apply for (covariant) derivatives of
$K^i_{ab}$, as these would appear at higher orders. Further, recall that the
entangling surface $\tilde\Sigma$ on the brane is defined by the intersection
of the extremal surface with $\rho=1$ and hence eq.~\reef{inducemet} yields
 \be
\hh_{ab}= h_{ab}\big|_{\rho=1}=\hz_{ab}+\ho_{ab}+\cdots
 \labell{bondx}
 \ee
for the induced metric on the $\tilde\Sigma$. Again the curvature constraints,
\reef{cond1} and \reef{cond2}, ensure that the differences between these two
metrics is small, \ie using eq.~\reef{indmetric}, we have
 \be
\hh_{\,ij}\,-\hz_{ij}\sim \ho_{ij}\ll1\,.
 \labell{okax}
 \ee

The discussion up to this point was absolutely general, and there was no need
to specify the details of the bulk gravity action in the bulk. However, the
detailed expressions for the holographic entanglement entropy across
$\tilde\Sigma$ are sensitive to the form of this action. Next, we illustrate
this calculation using the usual prescription \reef{define} for the case where
the bulk theory is just Einstein gravity (coupled to a negative cosmological
constant). Then we follow with a brief discussion describing results for
Gauss-Bonnet gravity in the bulk. In this case, we use the generalized
prescription of \cite{gbee,jan} to calculate the holographic entanglement
entropy.

\subsection{Einstein gravity}

Our bulk gravity action consists of the Einstein-Hilbert action with a negative
cosmological constant and we must also include the usual
Gibbons-Hawking surface term\footnote{Calligraphic $\R$ and $\K$ will be used
to denote bulk curvature and the second fundamental form of the brane
respectively. We implicitly assume that bulk integral runs over both copies of
the AdS space whereas surface integral is carried over both sides of the
brane.}
 \be
I_{bulk}^\mt{E}=\frac {1}{16\pi G_{d+1}}\int d^{d+1}x \sqrt{-G}
\Big[\frac{d(d-1)}{\delta^2}+\R \Big] +{1\over 8\pi G_{d+1}}\int d^{d}x
\sqrt{-\tg} \, \K ~.
 \labell{EHact}
 \ee
In appendix \ref{effact}, we showed that with this bulk theory, the induced
gravity action on the brane is given by
 \be
I^\mt{E}_{ind}=\int d^{d}x\sqrt{-\tg} \[ {R\over 16\pi G_{d}}
+\,\frac{\kappa_1}{2\pi} \bigg(R_{ij}R^{ij}-{d\over 4(d-1)}R^2\bigg)
+\mathcal{O}(\del^6)
\] ~,
 \labell{EHind0}
 \ee
where the expressions defining the effective Newton's constant and the
curvature-squared coupling in terms of $\delta$ and $G_{d+1}$ or the central
charge are given in eqs.~\reef{enewton} and \reef{ekappa1}.

The holographic entanglement entropy for generic entangling surfaces in the
boundary is evaluated using eq.~\reef{define}. We begin by evaluating the area
$\A(\sigma)$ of the extremal surface to the first two leading orders in the
expansion given in eq.~\reef{inducemet}\footnote{Factor two accounts for the
two copies of AdS space in the construction.}
 \bea
\A(\sigma)&=&2\int d^{d-2}y\, d\rho\, \sqrt{h}
 \labell{arealx}\\
 &=&\int_{\tilde\Sigma}  d^{d-2}y
\int_1^\infty d\rho\, \frac{\delta}{\rho^{d/2}}\,\sqrt{ \hz}\left[1
+\left(\ho_{\rho\rho} + \hz^{ab}\,\ho_{ab}
\right)\frac{\rho}2+\mathcal{O}(\del^4)\right]\,.
 \nonumber
 \eea
Now we can use eq.~\reef{bondx} to re-express this result in terms of induced
metric on the brane $\hh_{ab}$ rather than the boundary metric
$\overset{\scriptscriptstyle{(0)}}{h}_{ab}$. In particular, we have
 \be
\sqrt{ \hz}=\sqrt{  \hh}\(1-{1\over2}
\hz^{ab}\ho_{ab}+\mathcal{O}(\del^4)\)~.
 \labell{measure}
 \ee
Recall that the difference between the two metrics is small, as shown in
eq.~\reef{okax}. Therefore explicitly applying the conversion to $\hh_{ab}$ in
the first-order terms here and in eq.~\reef{arealx} is not necessary. This
would only generate terms of order $O(\del^4)$, which we are not evaluating
here. Now carrying out integration over $\rho$ in eq.~\reef{arealx} (and
keeping only the lower limit at $\rho=1$) yields
 \be
S_\mt{EE}={\delta\over 2(d-2)G_{d+1}}\int_{\tilde\Sigma} d^{d-2}y\sqrt{
\hh}\left[1 +{d-2\over 2(d-4)}\ho_{\rho\rho} +{1\over d-4}\hz^{ab}\,\ho_{ab}
+\mathcal{O}(\del^4)\right] \, .
 \labell{EEEE4}
 \ee
Finally we can substitute for $\overset{\scriptscriptstyle{(1)}}{h}_{\al\beta}$
using eq.~\reef{indmetric} and at the same time, we use eqs.~\reef{enewton} and
\reef{ekappa1} to express the result in terms of the gravitational couplings in
the induced action \reef{EHind0}. Our final expression for the entanglement
entropy becomes
 \be
S_\mt{EE}=\frac{{\cal A}(\tilde\Sigma)}{4G_d} +
\kappa_1\,\int_{\tilde\Sigma}d^{d-2}y\sqrt{  \hh} \(
2R^{ij}\,{\tg}^{\perp}_{ij}-{d\over d-1}\,R- K^iK_i \)+\mathcal{O}(\del^4) \, .
 \labell{EHent}
 \ee
Here, all curvatures are evaluated on the entangling surface $\tilde\Sigma$ and
${\tg}^{\perp}_{ij} =\eta_{\hi\hj}\,n^\hi_i\,n^\hj_j$ is the metric in the
transverse space to the entangling surface, \ie
${\tg}^{\perp}_{ij}=\tg_{ij}-\hh_{ij}$.

The first important feature to note about this result is that leading term
precisely matches the BH formula \reef{prop0} for the induced gravity theory
\reef{EHind0}. However, here it appears in $S_\mt{EE}$ for a general
entangling surface rather than a horizon entropy. That is, subject to the
constraints in eqs.~\reef{cond1} and \reef{cond2} in this RS2 model, we find
that the leading contribution to the entanglement entropy for any general
(large) regions is given precisely by the Bekenstein-Hawking formula. Of
course, this result precisely matches the conjecture of \cite{new1}!

The next-to-leading term in eq.~\reef{EHent} reveals a non-trivial correction
to the area law. The appearance of $\kappa_1$ here suggests that it is
connected to the curvature-squared interaction appearing in the induced gravity
action \reef{EHind0}. Of course, this connection naturally brings to mind the
Wald entropy \cite{WaldEnt}, which describes the horizon entropy of
(stationary) black hole solutions in theories with higher curvature
interactions. In particular, let $\tS$ be (a cross-section of) a Killing
horizon in a gravity theory with a general (covariant) Lagrangian $\mathcal{
L}(g,R,\nabla R,\cdots)$. Then the Wald entropy is \cite{WaldEnt}
 \be
S_\mt{Wald}=-2\pi\int_{\tS}d^{d-2}y\sqrt{  \hh}\ {\del \mathcal{ L}
\over\del R^{ij}{}_{kl}} \,\hat\veps^{ij}\,\hat\veps_{kl}~,
 \labell{wald}
 \ee
where as above, $\hh_{ab}$ is the induced metric on $\tS$ and $\hat\veps_{ij}$
is the volume-form in the two-dimensional transverse space to $\tS$. Some
useful identities for the latter include:\footnote{Recall that the signature of
the transverse space is $(-,+)$.}
 \be
\hat\veps_{ij}\,\hat\veps_{kl}= {\tg}^{\perp}_{il}\,{\tg}^{\perp}_{jk}
-{\tg}^{\perp}_{ik}\, {\tg}^{\perp}_{jl}\, ,\qquad
\hat\veps_{ik}\,\hat\veps_{j}{}^k= -\,{\tg}^{\perp}_{ij}\, , \quad
\hat\veps_{ij}\,\hat\veps^{ij}=-2\,. \labell{useful}
 \ee
Applying eq.~\reef{wald} (as well as the above identities) to the induced
gravity theory \reef{EHind0}, we obtain
 \be
S_\mt{Wald}=\frac{{\cal A}(\tilde\Sigma)}{4G_d} +
\kappa_1\,\int_{\tilde\Sigma}d^{d-2}y\sqrt{  \hh} \(
2R^{ij}\,{\tg}^{\perp}_{ij}-{d\over d-1}\,R \)+\mathcal{O}(\del^4) \, .
 \labell{wald2}
 \ee
Comparing eqs.~\reef{EHent} and \reef{wald2}, we see that $S_\mt{EE}$ and
$S_\mt{Wald}$ agree up to the absence of the extrinsic curvature terms in the
Wald entropy. However, this discrepancy might have been expected since, as we
emphasized above, the Wald formula \reef{wald} was constructed to be applied to
Killing horizons, for which the extrinsic curvature vanishes.\footnote{On a
Killing horizon, the extrinsic curvature will vanish precisely on the
bifurcation surface. For a general cross-section of the Killing horizon, the
extrinsic curvature is nonvanishing but only for a null normal vector. Hence
one finds that any scalar invariants constructed with the extrinsic curvature
still vanish, \eg in general, $K^i\ne0$ however $K^iK_i=0$.} Hence if
eq.~\reef{EHent} is evaluated on a Killing horizon, we will find $S_\mt{EE}=
S_\mt{Wald}$.

\subsection{Gauss-Bonnet gravity} \labell{GBtangle}

In this section we analyze higher curvature gravity in the bulk. Our discussion
will focus on Gauss-Bonnet (GB) gravity, and we regard the latter as simply a
convenient toy model which may provide some insights into more general bulk
theories. The bulk action is given by
 \be
I_{bulk}^\mt{GB}=\frac {1}{16\pi G_{d+1}}\int d^{d+1}\!x\, \sqrt{-G}
\left[\frac{d(d-1)}{L^2}+\R+\frac{{L}^2\,\lambda}{(d-2)(d-3)}\,\chi_4
\right]+I^\mt{GB}_{surf}\,.
 \labell{GBact}
 \ee
where $\chi_4$ is proportional to the four-dimensional Euler density,
 \beq
\chi_4=\R_{\mu\nu\rho\sigma}\R^{\mu\nu\rho\sigma}-4\,
\R_{\mu\nu}\R^{\rho\sigma}+\R^2\,. \labell{euler4}
 \eeq
The detailed form of the surface term $I^\mt{GB}_{surf}$ is given in
eq.~\reef{GBsurf}. Now with the above bulk action, we showed in appendix
\ref{effact} that the induced gravity action for the RS2 braneworld becomes
 \be
I^\mt{GB}_{ind}=\int d^{d}x\sqrt{-\tg} \[ {R\over 16\pi  G_{d}}
+\,\frac{\kappa_1}{2\pi}\bigg(R_{ij}R^{ij}-{d\over 4(d-1)}R^2\bigg)
+\frac{\kappa_2}{2\pi} \,C_{ijkl} C^{ijkl}+\mathcal{O}(\del^6)  \] ~.
  \labell{EGBind0}
 \ee
where $C_{ijkl}$ is the Weyl tensor of the brane geometry. The $d$-dimensional
Newton's constant and the couplings for the curvature-squared terms are defined
in eqs.~(\ref{eegb2x}--\ref{kappa2}).

The prescription for the holographic entanglement entropy is modified for GB
gravity \cite{gbee,jan}. In particular, it still involves extremizing over bulk
surfaces as in the original prescription \reef{define} but the functional to be
evaluated on these surfaces is no longer the BH formula. Rather the latter is
replaced by the following expression:
 \be
 S_{\mt{JM}}=\frac {1}{2 G_{d+1}}\int_\sigma d^{d-2}y\, d\rho \sqrt{h}
 \[ 1+\frac{2\,L^2\,\lambda}{(d-2)(d-3)}\,\mathcal{R} \]
 +\frac{2\, L^2\,\lambda}{(d-2)(d-3)G_{d+1}}\int_{\tS}\mathcal{K} \, ,
 \labell{eegb}
 \ee
where $\mathcal{R}$ is intrinsic curvature of the bulk surface $\sigma$,
$\mathcal{K}$ is the trace of the second fundamental form on the boundary of
$\sigma$, which coincides with the entangling surface $\tS$ on the brane. In
eq.~\reef{eegb}, we already introduced a factor two to account for both copies
of AdS space on either side of the brane. Apart from this factor of two, we
note that $S_\mt{JM}$ was derived to describe black hole entropy in GB gravity
\cite{ted0} but it only coincides with $S_\mt{Wald}$ for surfaces with
vanishing extrinsic curvature \cite{gbee}.

As before, we assume that the background geometry on the brane and the
entangling surface $\tS$ are big enough such that eqs.~\reef{cond1} and
\reef{cond2} are satisfied. Then derivative expansion can be applied to make a
Taylor series expansion of the intrinsic and extrinsic curvatures, $\R$ and
$\K$, however, we relegate details to appendix \ref{geom}. Substituting
eqs.~\reef{Rsigma1} and \reef{Ksigma} into eq.~\reef{eegb} and integrating out
radial direction $\rho$, yields
 \bea
S_\mt{EE}&=&\frac {{\cal A}(\tS)}{4 G_{d}} +\kappa_1\int_{\tS} d^{d-2}y
\sqrt{  \hh} \[ 2R^{ij}\, {\tg}^{\perp}_{ij}-{d\over d-1}\,R-K^iK_i
\]
\labell{entJM}\\
&&\quad +\ 4\kappa_2\int_{\tS}d^{d-2}y \sqrt{  \hh}
\[\hh^{ac}\hh^{bd}C_{abcd}-K^i_{ab}K_i{}^{ab}+\frac{1}{d-2}K^iK_i\]
+\mathcal{O}(\del^4) \, . \nonumber
 \eea

Again, we find that in this RS2 model, the leading contribution to the
entanglement entropy evaluated for arbitrary large regions is given precisely
by the Bekenstein-Hawking formula \reef{prop0}, in agreement with the
conjecture of \cite{new1}. As in the previous section, we can also compare
above result with the Wald entropy \reef{wald} for the induced gravitational
action \reef{EGBind0}. Again $S_\mt{EE}$ and $S_\mt{Wald}$ match except that
the extrinsic curvature terms above do not appear in the Wald entropy.

As a final note, it is amusing to observe that the geometric terms appearing in
eq.~\reef{entJM} are almost the same. Using the geometric identities provided
in appendix \ref{geom}, we can write
 \bea
&&2R^{ij}\, {\tg}^{\perp}_{ij}-{d\over d-1}\,R-K^iK_i = \labell{corn}\\
&&\qquad\qquad\qquad {d-2\over d-3} \[
\hh^{ac}\hh^{bd}C_{abcd}-K^i_{ab}K_i{}^{ab}+\frac{1}{d-2}K^iK_i - R_{\tS}\]\, ,
 \nonumber
 \eea
where $R_{\tS}$ denotes the intrinsic Ricci scalar of the entangling surface
$\tS$. Given this expression, eq.~\reef{entJM} can be rewritten as
 \bea
S_\mt{EE}&=&\frac {{\cal A}(\tS)}{4 G_{d}} -{d-2\over
d-3}\,\kappa_1\int_{\tilde\Sigma}d^{d-2}y \sqrt{  \hh}\  R_{\tS}
 \labell{entJM2}\\
&&\qquad\qquad+ \kappa_3\int_{\tilde\Sigma}d^{d-2}y \sqrt{  \hh}
 \[\hh^{ac}\hh^{bd}C_{abcd}-K^i_{ab}K_i{}^{ab}+\frac{1}{d-2}K^iK_i\]
+\mathcal{O}(\del^4) \, ,
 \nonumber
 \eea
where
 \be
\kappa_3 = 4\,\kappa_2+{d-2\over d-3}\,\kappa_1={2\over
\pi(d-2)(d-3)(d-4)}\,{\ct\over \delta^{d-4}}\,.
 \labell{kappa3}
 \ee
The last expression for the new coupling $\kappa_3$ comes from combining
eqs.~\reef{kappa1} and \reef{kappa2}. Now it is interesting to consider this
result in the special case $d=4$. In this case, the $\kappa_n$ couplings are
all dimensionless, but at the same time the expressions that we have provided
above and in appendix \ref{effact} are not quite correct --- they all appear to
diverge because of a factor $1/(d-4)$. Re-visiting the derivation of these
expressions, one finds that in fact these couplings contain a logarithmic
dependence on the cut-off $\delta$. In particular, we write for $d=4$:
 \be
\kappa_1 = -{\ads\over 2\pi}\log(\mu\delta)\,, \qquad \kappa_3 = -{\ct\over
\pi}\log(\mu\delta)\,.
 \labell{kappa134}
 \ee
where $\mu$ is some renormalization scale. Further note that with the
normalization chosen in eqs.~\reef{effectc} and \reef{effecta}, the central
charges, $\ct$ and $\ads$ match precisely the standard central charges
appearing in the trace anomaly, \ie $\ads=a$ and $\ct=c$ \cite{cthem,cc}.
Hence, the entanglement entropy \reef{entJM2} becomes, for $d=4$
 \bea
S_\mt{EE}&=&\frac {{\cal A}(\tS)}{4\tilde G_{4}}
 \labell{donkey}\\
&&\quad- {\log(\mu\delta)\over\pi}\int_{\tilde\Sigma}d^{d-2}y \sqrt{
\hh}\(c
 \[\hh^{ac}\hh^{bd}C_{abcd}-K^i_{ab}K_i{}^{ab}+\frac{1}{d-2}K^iK_i\]-a\,
 R_{\tS}\)
+\cdots \, .
 \nonumber
 \eea
We can recognize the second term above as the universal contribution to the
entanglement entropy of a four-dimensional CFT \cite{solo}. Actually, the
attentive reader may notice that there is an extra overall factor of two, which
arises because there are actually two copies of the CFT corresponding to the
two copies of AdS space.


\section{Beyond the Area} \labell{test}

Recent progress has revealed an interesting interplay between entanglement
entropy and renormalization group flows, \eg \cite{cthem,two,three,maze,flow}.
One important result is an elegant proof for the c-theorem in two dimensions
\cite{zamo} formulated in terms of entanglement entropy \cite{two}. In
particular, one begins by considering the entanglement entropy on an interval
of length $\ell$ and then evaluates
 \be
 C_2(\ell)\equiv \ell\,\partial_\ell S(\ell)\,.
 \labell{c2}
 \ee
If the underlying field theory is a two-dimensional CFT, then $C_2$ is a
constant independent of $\ell$ and in fact, $3\,C_2=c$, the central charge
characterizing the CFT. Now in general, if one assumes only Lorentz invariance,
unitarity and strong subadditivity \cite{Lieb}, one can demonstrate \cite{two}
 \be
\partial_\ell C_2(\ell)\le 0\,.
 \labell{ineq2}
 \ee
Therefore comparing $C_2$ found at short scales with that determined by probing
the system at long distances, one has $\[C_2\]_{UV}\ge \[C_2\]_{IR}$ and of
course, if the underlying field theory describes an RG flow connecting two
fixed points, then the same inequality holds for the corresponding central
charges. In an exciting recent development, \cite{three} extended this
construction to prove an analogous c-theorem which had been conjectured for
three dimensions \cite{cthem,Fthem}. In three dimensions, one considers the
entanglement entropy of a disk of radius $R$ and arrives at the following
construction \cite{three,maze}
 \be
 C_3(R)\equiv R\,\partial_R S(R)-S(R)\,,
 \labell{c3}
 \ee
which yields an interesting (constant) central charge in the case where the
underlying theory is a CFT. In general, again with the assumptions of Lorentz
invariance, unitarity and strong subadditivity, one can establish the following
inequality:
 \be
\partial_R C_3(R)=R\, \partial^2_R S\le 0\,,
 \labell{ineq3}
 \ee
which establishes the three-dimensional version of the c-theorem.

Now, turning to higher dimensions, one can observe \cite{flow,tadashi} the
inequality \reef{ineq2} will still apply in any situation where the background
geometry preserves Lorentz symmetry in a plane and the entangling surface is
chosen as two points (spacelike) separated in this plane by a distance $\ell$.
The simplest example to consider is a `strip' or `slab' geometry in $R^d$, \ie
the entangling surface is chosen to be two parallel $(d-2)$-dimensional planes
separated by a distance $\ell$ along the $x$-axis
--- see figure \ref{python}a. As before, one can evaluate the entanglement
entropy for the region between the two planes and then construct the function
$C_2(\ell)$, as in eq.~\reef{c2}. However, note that $C_2(\ell)$ will not be a
constant even when the underlying theory is a CFT for $d\ge3$ \cite{flow}. The
geometric approach of \cite{two} only relies on making Lorentz transformations
in the $(t,x)$-plane and then comparing entropies for different pairs of
planes. Hence with the same assumptions of Lorentz invariance, unitarity and
strong subadditivity, the inequality \reef{ineq2} again holds in this
situation.
\FIGURE[!ht]{
 \begin{tabular}{ccc}
\includegraphics[width=0.5\textwidth]{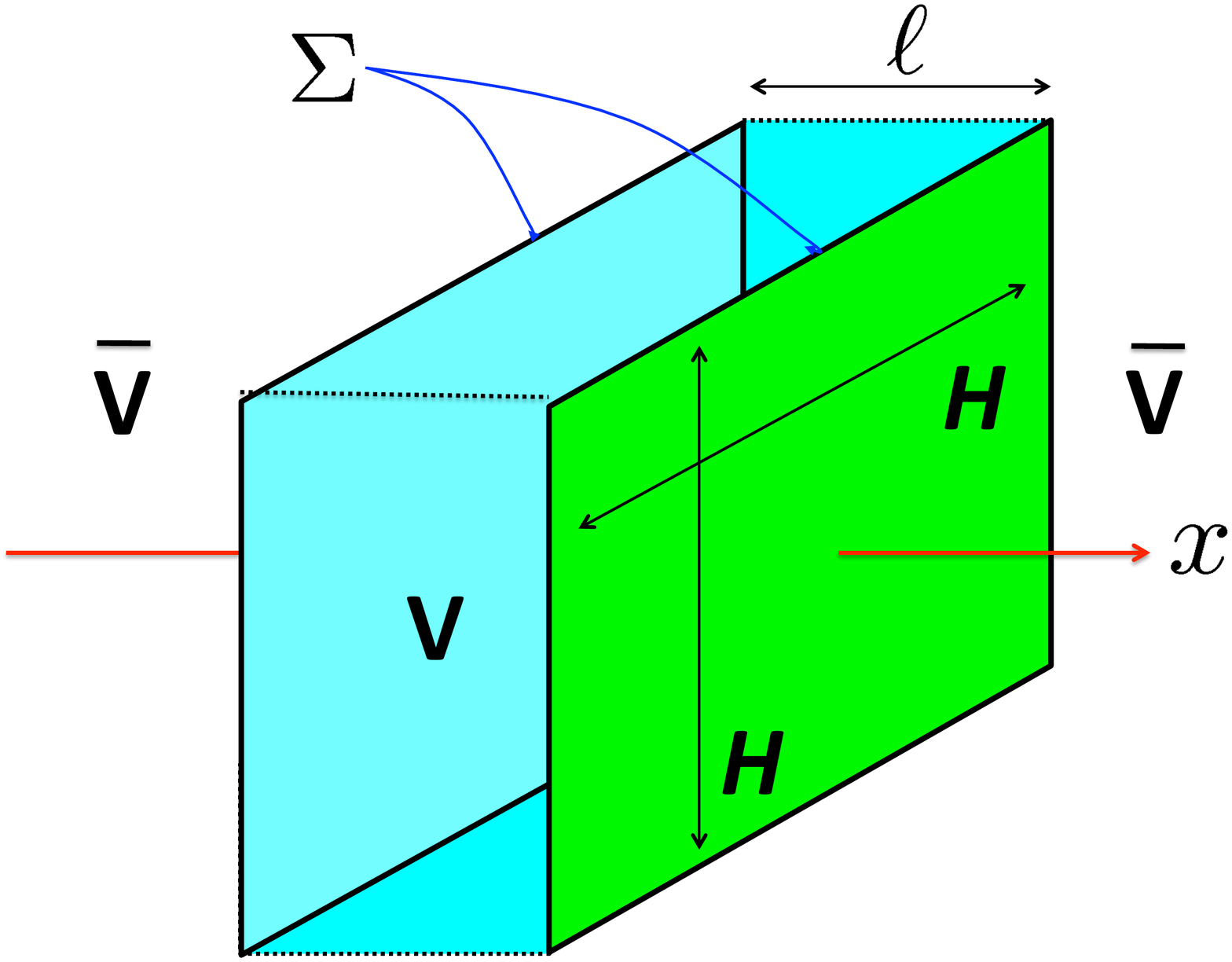}&&
\includegraphics[width=0.57\textwidth]{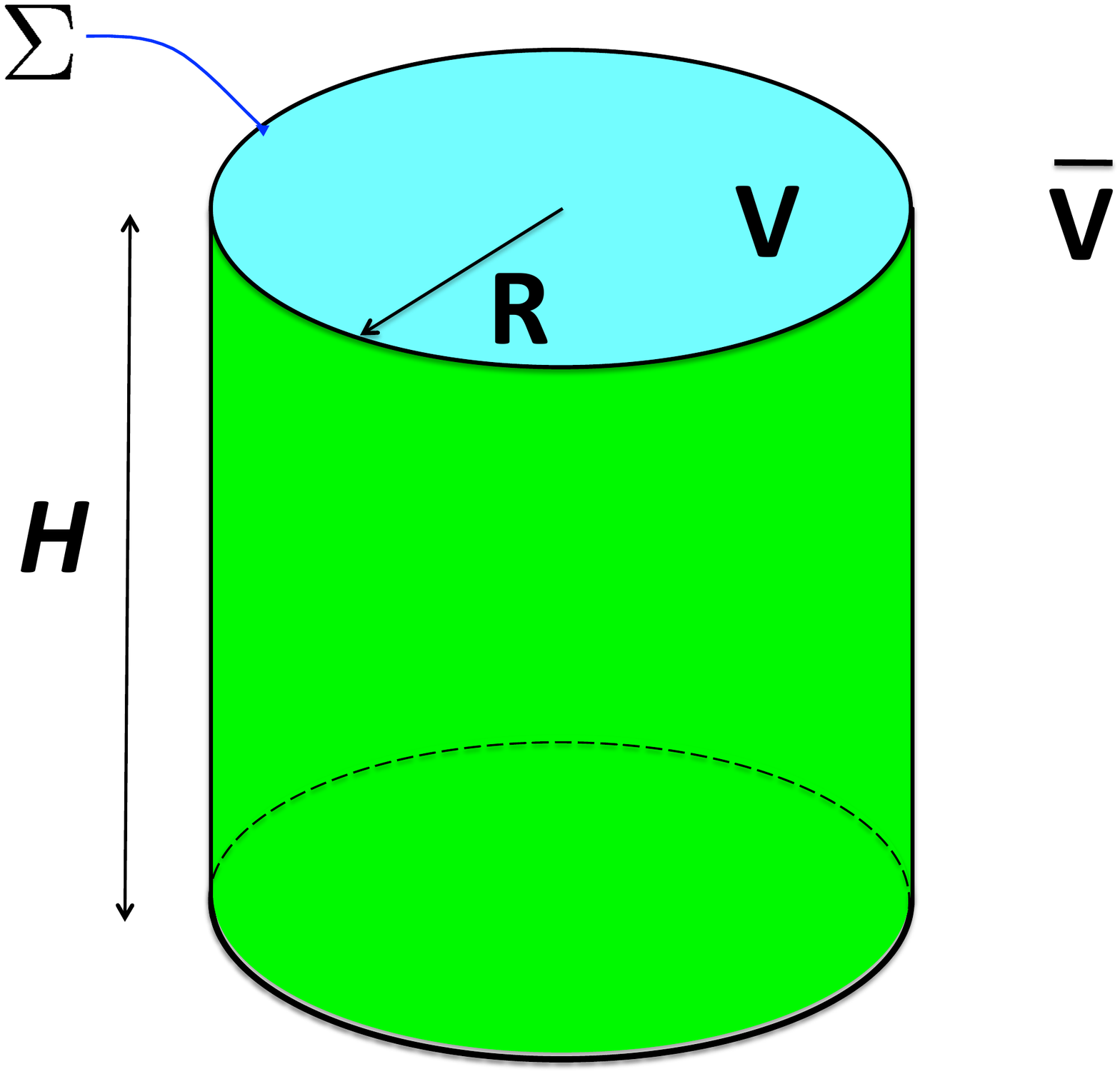}\\
(a) & & (b)
\end{tabular}
\caption{(Colour Online) Panel (a) shows the slab geometry on a constant time
slice. The entangling surface consists of two parallel (hyper)planes separated
by a distance $\ell$. The reduced density matrix is calculated for the region
$V$ between these two planes by integrating out the degrees of freedom in the
exterior region $\bar V$. Panel (b) shows a cylindrical entangling geometry
with radius $R$. In both cases, the distance $H$ is introduced to regulate the
area of the entangling surfaces.}
 \label{python}}

Similarly, the inequality \reef{ineq3} will apply in higher dimensions, as long
as the background geometry preserves Lorentz symmetry in a three-dimensional
Minkowski subspace and the entangling surface is chosen as a circle in a
spacelike plane in this subspace (without any additional structure in the extra
dimensions). Of course, the simplest example to consider is a cylindrical
entangling surface in $R^d$, \ie the ($d-2$)-dimensional entangling surface has
topology $S^1\times R^{d-3}$, as shown in figure \ref{python}b. Here the
approach of \cite{three} can again be applied to establish the inequality
\reef{ineq3} for $C_3(R)$, which is again constructed as in eq.~\reef{c3}.

In the following, we will consider testing our holographic results for the RS2
model with the above inequalities, \reef{ineq2} and \reef{ineq3}. In this case,
the bulk geometry will still be empty AdS space and so we are not considering a
nontrivial RG flow in the boundary CFT. However, in comparison to
\cite{two,three}, there are unconventional aspects of the present calculations,
including that the underlying degrees of freedom include gravity and that the
boundary CFT has an explicit cut-off $\delta$. On the other hand, it seems that
the basic assumptions of \cite{two,three} still seem to apply in the present
context, \ie Lorentz invariance, unitarity and strong subadditivity. Hence we
will find that demanding that our results for slab and cylindrical geometries
satisfy eqs.~\reef{ineq2} and \reef{ineq3}, respectively, provide new insights
into our model. For simplicity, we will only present our calculations for the
case with Einstein gravity in the bulk.

\subsection{Slab geometries}

We begin by considering the slab geometry shown in figure \ref{python}a for
$d\ge3$. We will denote the separation of the two planes on the brane as $\tl$
and reserve $\ell$ to denote the corresponding distance on the AdS boundary in
our holographic calculations. Note that from our previous calculations, we can
expect that the BH term \reef{prop0} will appear as the leading contribution in
the entanglement entropy, \ie
 \be
S_\mt{EE}= \frac{H^{d-2}}{2G_d}+\cdots\,,
 \labell{wallS}
 \ee
where $H^{d-2}$ corresponds to the regulated area of one of the planes and
hence the total area of the entangling surface is $\A(\tS)=2H^{d-2}$. Note that
this leading term is independent of the separation $\tl$ and so $C_2(\tl)$
depends entirely on the higher order terms in eq.~\reef{wallS}. Further, since
the background geometry is flat space and the entangling surface itself is
flat, any higher order geometric contributions, like those explicitly shown in
eq.~\reef{entJM2}, will vanish. Hence the contributions that we are probing in
our calculations here should be thought of as coming from long-range
correlations in the CFT. From previous holographic calculations \cite{flow}, we
can expect that to leading order, $C_2(\tl)$ takes the form
 \be
C_2(\tl)=\pi\,\gamma^{d-1}\, \ct\, \frac{H^{d-2}}{\tl^{d-2}} +\cdots\,, \quad
{\rm with}\ \ \gamma=\frac{\Gamma(\frac{1}{2(d-1)})}{2\sqrt{\pi}\,
\Gamma(\frac{d}{2(d-1)})}\,.
 \labell{c22}
 \ee

As the corresponding holographic calculations have been extensively described
elsewhere, \eg \cite{rt1,flow}, our description here is brief. To begin, we
write the AdS metric in Poincar\'e coordindates
 \be
 ds^2_{d+1}=\frac{\delta^2}{z^2}\left(-dt^2+dx^2+d\vec{y}^2 +dz^2\right)\,.
 \ee
where $y^i$ with $i=1$, 3, $\cdots$, $d-2$ describe the directions parallel to
the entangling surface. In the standard holographic calculation, one sets the
planes defining the entangling surface at $x=\ell/2$ and $x=-\ell/2$ where
$\ell$ denotes the separation at the AdS boundary $z=0$. As above, we set the
area of each of the two planes to be $H^{d-2}$, where $H$ is an arbitrary IR
regulator with $H\gg\ell$. As usual, the entanglement entropy is evaluated with
eq.~\reef{define} and area is extremized by a bulk surface with a profile
$x(z)$ satisfying
 \be
x' = \frac{z^{d-1}}{\sqrt{(\gamma \ell)^{2(d-1)}-z^{2(d-1)}}}\,.
 \labell{profile}
 \ee
For $d\geq3$, the final result can be written as
 \be
S_\mt{EE}=\frac{H^{d-2}}{2G_d}\left[ _2F_1\left(\frac{2-d}{2(d-1)},
\frac{1}{2}, \frac{d}{2(d-1)},
\left(\frac{\delta}{\gamma\ell}\right)^{2(d-1)}\right)
 -\frac{1}{2\gamma}\left(\frac{\delta}{\gamma\ell}\right)^{d-2}\right]\, ,
 \labell{answer1}
 \ee
where the effective $d$-dimensional Newton's constant is given by
eq.~(\ref{newton}). If this expression is expanded for $\delta\ll\ell$, we
recover the expected area law, as in eq.~\reef{wallS}. Now this result is
written in terms of $\ell$, the separation of the two planes on the AdS
boundary, whereas we would like to express the results in terms of $\tl$, the
separation on the brane. The relation between these two distances is readily
found by integrating eq.~\reef{profile} between $z=0$ and $z=\delta$, with the
final result given by
 \be
 \tl=\ell\left[1-\frac{2}{d}\left(\frac{\delta}{\gamma \ell}\right)^{d}\,
  _2F_1\left(\frac{1}{2}, \frac{d}{2(d-1)}, \frac{3d-2}{2(d-1)},
  \left(\frac{\delta}{\gamma\ell}\right)^{2(d-1)}\right)\right] \, .
  \labell{length1}
  \ee

\begin{figure}[t]
\centering
{\includegraphics[width=200pt,height=160pt]{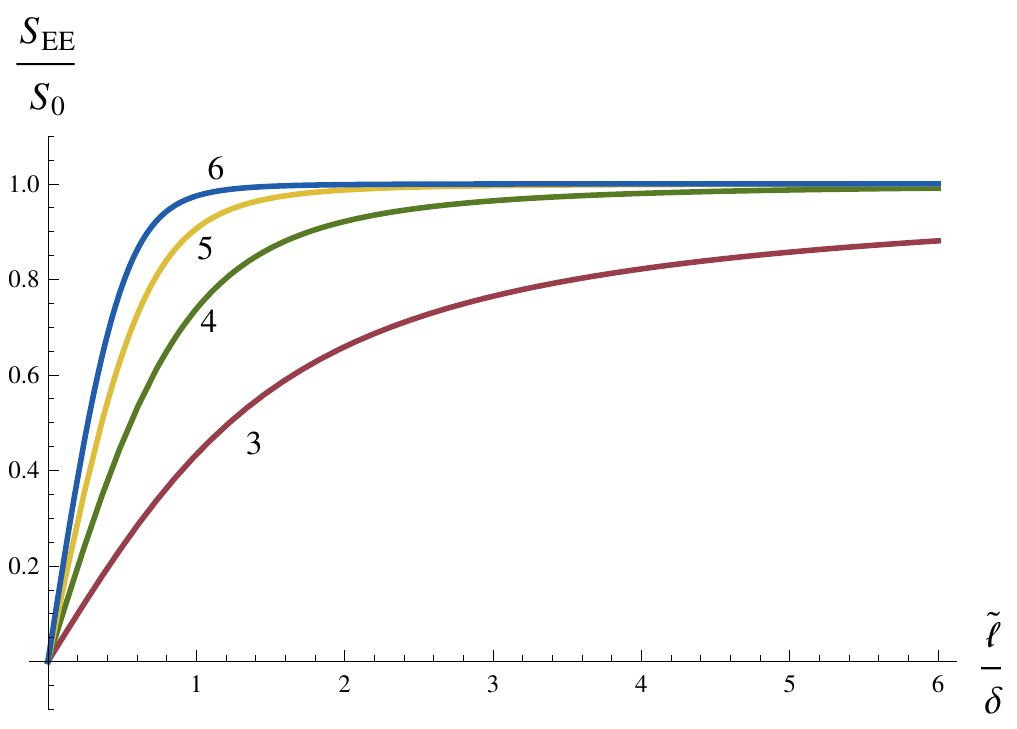}}
\centering
{\includegraphics[width=200pt,height=160pt]{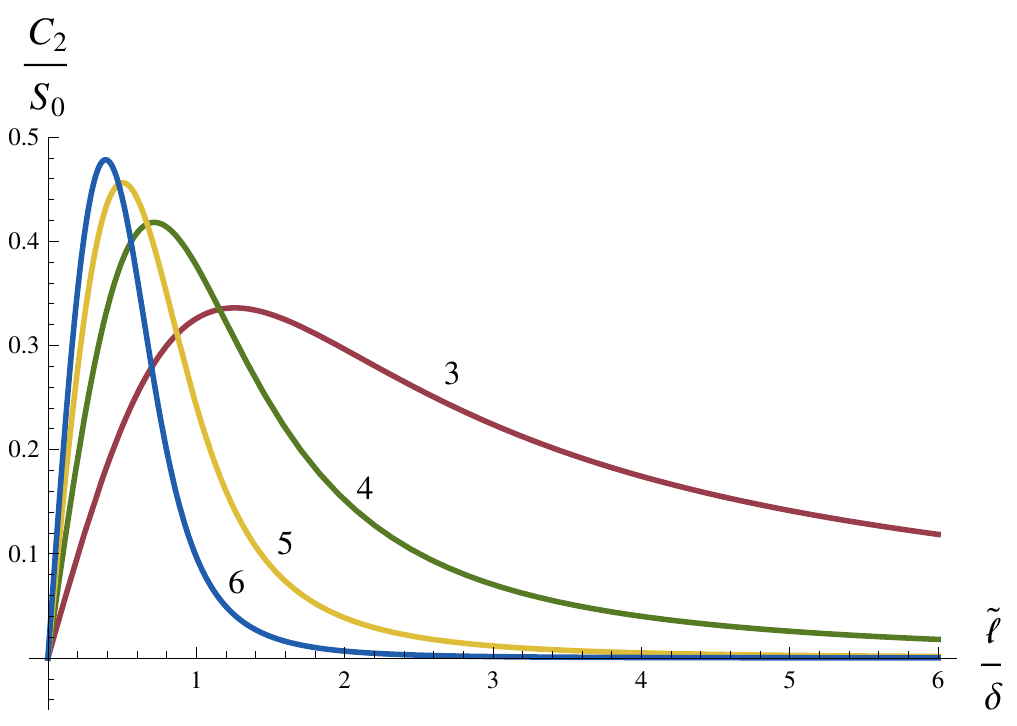}}
\centering
{\includegraphics[width=210pt,height=170pt]{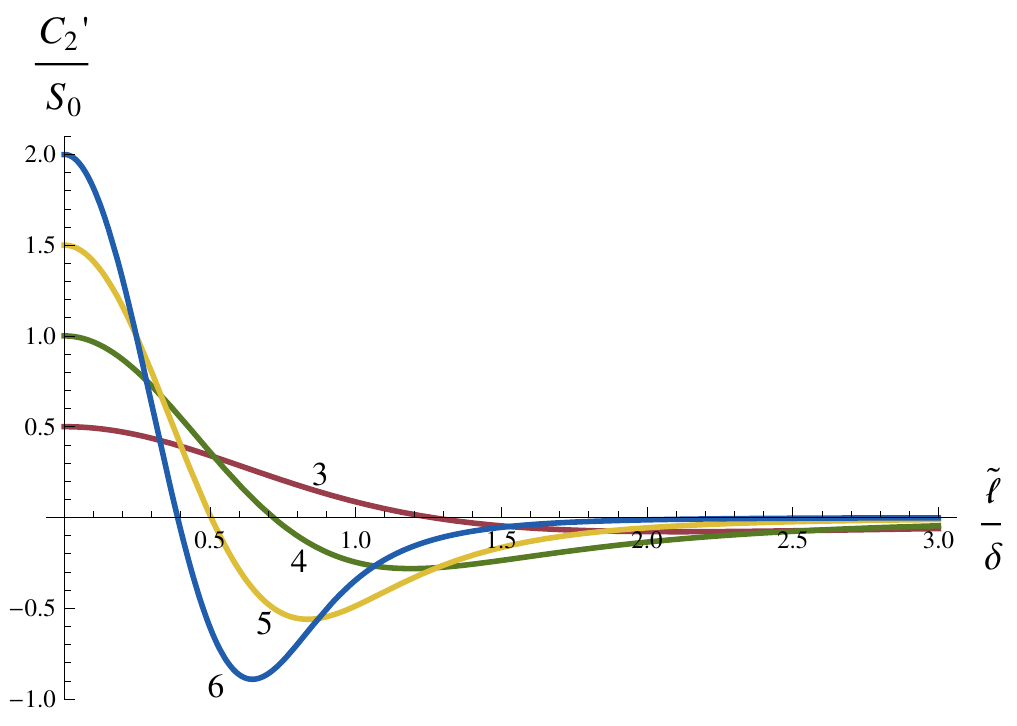}}
\caption{(Colour online) $S_\mt{EE}$, $C_2$ and $C'_2=\delta\partial_{\tl}C_2$
as a function of $\tl$ for $d=3, 4, 5, 6$. The vertical axes are normalized
with $S_0=\frac{H^{d-2}}{2G_d}$. The first plot confirms that for
$\tl\gg\delta$, the dominant contribution in entanglement entropy is
the BH term, \ie $S_0$.
Also the last plot reveals that $C'_2$ becomes positive for
$\tilde{\ell}\lesssim\delta$, indicating a limitation with this model.
}\label{fig:d}
\end{figure}
Given eqs.~\reef{answer1} and \reef{length1}, figure \ref{fig:d} plots the
results for $S_\mt{EE}$, $C_2$ and $\partial_{\tl}C_2$ in terms of
$\tl/\delta$, for $d\geq3$. The plot of the entanglement entropy confirms that
$S_\mt{EE}\to S_0=H^{d-2}/(2G_d)$ asymptotically for $\tl/\delta\to\infty$ but
note that $S_\mt{EE}- S_0<0$ for all values of $\tl$. Further $S_\mt{EE}$ goes
to zero at $\tl=0$, as would be expected since the region $V$ has shrunk to
zero size at this point. Now the plot of $C_2(\tl)$ shows that it is increasing
for relatively small separations, \ie $\tl\lesssim\delta$, and it decreases for
large values of $\tl$. Hence in the next plot, we see $\partial_{\tl}C_2$ is
negative as required when the separation is large. However, we also find
$\partial_{\tl}C_2>0$ for $\tl\lesssim\delta$.

Presumably we have found an inconsistency in our model for small
separations, \ie $\tl\sim\delta$. Of course, it should not be surprising to
find unusual behaviour when the width of the slab is of the same order as the
short-distance cut-off. In particular, with this intrinsic cut-off, the model
has only a finite resolution of order $\delta$ and hence it is not actually
meaningful to consider evaluating the entanglement entropy for the slab when
$\tl\lesssim\delta$. Essentially the assumption of strong subadditivity is lost
at this scale because we cannot effectively distinguish the degrees of freedom
inside and outside of the slab. The fact that $\partial_{\tl}C_2$ becomes
positive in this regime is simply pointing out this limitation of the model.

\subsection{Cylindrical geometries}

In this section, we examine the entanglement entropy for a cylindrical
entangling surface with $d\ge3$, \ie $\tS=S^1\times R^{d-3}$ in a flat $R^d$
background, as shown in figure \ref{python}b. We will denote the radius of the
circle on the brane as $\tR$ while $R$ will be the corresponding radius on the
AdS boundary.  Eq.~\reef{entJM2} indicates that the leading contributions to
the entanglement entropy should take the form
 \be
S_\mt{EE}=
\frac{\pi\tR\,H^{d-3}}{2G_d}-2\pi\kappa_3 {d-3\over d-2}\frac{H^{d-3}}{\tR}+\cdots\,,
 \labell{cylinS}
 \ee
where $H$ is the scale which regulates the area of $\tS$, \ie
$\A(\tS)=2\pi\tR\,H^{d-3}$. Hence we expect that for large radius
($\tR\gg\delta$), the BH area term \reef{prop0} will be the leading
contribution to $S_\mt{EE}$. However, note that the construction of $C_3$ in
eq.~\reef{c3} is designed to precisely remove the area term for the cylindrical
geometry \cite{maze} and so to leading order, we expect
 \be
C_3= 4\pi\kappa_3{d-3\over d-2}\frac{H^{d-3}}{\tR}+\cdots\,.
 \labell{c32}
 \ee
Hence in this case, $C_3(\tR)$ contains geometric terms arising from
short-range correlations across the entangling surface, as well as nonlocal
contributions coming from long-range correlations in the CFT.

To begin the holographic calculation, we write the AdS metric in Poincar\'e
coordinates as,
 \be
 ds^2_{d+1}=\frac{\delta^2}{z^2}\left(-dt^2+dr^2+r^2d\phi^2+ d\vec{y}^2 +dz^2\right)\,.
 \ee
where $y^i$ with $i=1$, 3, $\cdots$, $d-3$ describe the directions parallel to
the entangling surface. In the standard holographic approach, one would define
the entangling surface with $r=R$ at the AdS boundary $z=0$. The entanglement
entropy is then evaluated with eq.~\reef{define} and we consider bulk surfaces
with a profile $r(z)$. The induced metric on such a bulk surface then becomes
 \be
ds^2_{d-1}=\frac{\delta^2}{z^2}\left[\left(1+ r'^2(z)\right)dz^2+r^2d\phi^2+
d\vec{y}^{\,2}\right]\,.
 \labell{induced6}
 \ee
Using eq.~(\ref{newton}), the entanglement entropy can then be written as
 \be
S=\frac{\A(\tilde{\Sigma})}{4G_d}\,\frac{(d-2)\delta^{d-2}}{\tilde{R}}
\int_{\delta}^{z_*}\frac{r \sqrt{1+r'^2}}{z^{d-1}}dz\,,
 \labell{actionarea3}
 \ee
where $z_*$ is the maximum value of $z$ where the surface reaches $r=0$ and
closes off in the bulk. The above functional can be used to derive an equation
of motion in order for the profile $r(z)$ to extremize the area:
\begin{equation}
rr''-\left(1+\frac{d-1}{z}rr'\right)(1+r'^2)=0\,.
\end{equation}
The latter must be solved subject to the boundary conditions $r(z=0)=R$ on the
AdS boundary and $r'=0$ at $r=0$ to ensure that the surface closes of smoothly
in the bulk. For $d=3$, one can obtain an analytic solution, since the
calculation is a special case of the analysis given in appendix \ref{sphere}
--- also, see below. For $d>4$ and $R\gg\delta$, we can find the expansion of $r(z)$ and hence of entanglement entropy \reef{actionarea3} in inverse powers of $R/\delta$. We checked that the leading and next-to-leading terms match eq. \reef{cylinS}, which was based on our general geometric formula \reef{entJM2}.

However, in general, we had to resort to numerical methods
to solve for the profile and the entanglement entropy. Further, one must
integrate the profile from $z=0$ to $z=\delta$ to determine the relation
between $R$ and $\tR$.

\begin{figure}[t]
\centering
{\includegraphics[width=200pt,height=160pt]{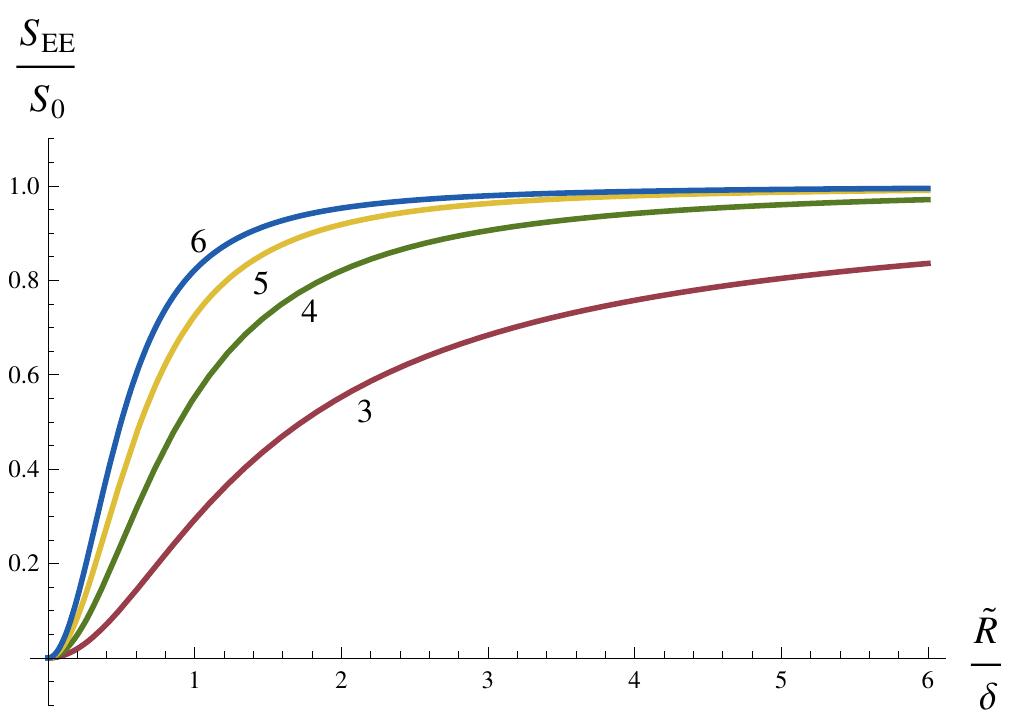}}
\centering {\includegraphics[width=200pt,height=160pt]{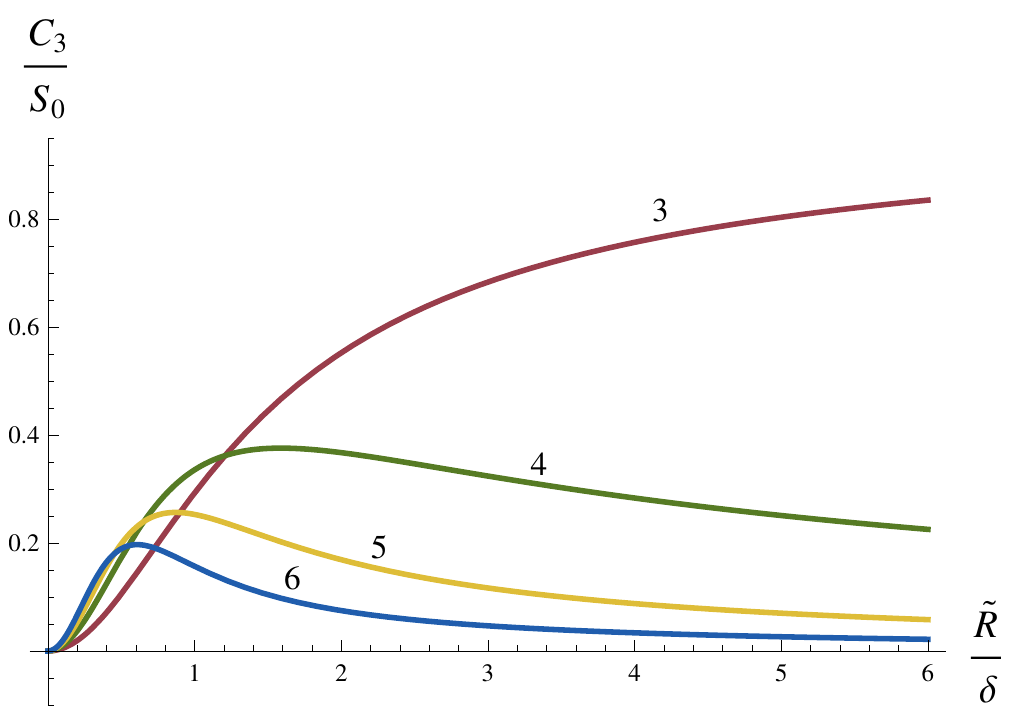}}
\centering
{\includegraphics[width=210pt,height=170pt]{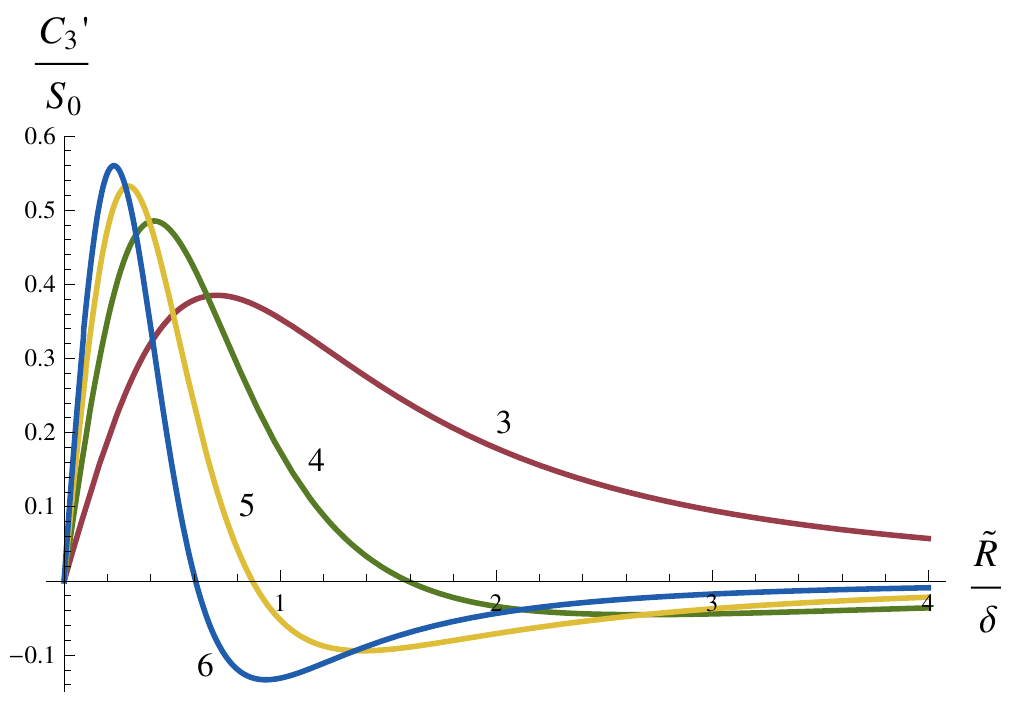}}
\caption{(Colour online)
$S_\mt{EE}$, $C_3$ and $C'_3=\delta\partial_{\tR}C_3$ as a function of $\tR$ for $d=3, 4,
5, 6$. The vertical axes are normalized with $S_0=\A(\tilde{\Sigma})/(4
G_d)$. The plot of $S_\mt{EE}$ confirms that for $\tR\gg\delta$, the dominant
contribution is the BH term, \ie $S_0$. The last plot reveals that for $d=4,
5, 6$, $C'_3$ becomes positive for $\tR\lesssim\delta$. Also note that for $d=3$,
$C'_3$ is positive for all $\tR$.}\label{fig:cyl}
\end{figure}
Figure \ref{fig:cyl} shows plots of $S_{EE}$, $C_3$ and its derivative as
functions of $\tR$ for $d=3$, 4, 5, and 6. The entropy plot confirms that
entanglement entropy is always positive and finite in terms of the radius of
the circle on the brane. It goes to zero at $\tilde{R}=0$, as expected since
the interior region shrinks to zero, and it is bounded from above by the
leading BH contribution shown in eq.~\reef{cylinS}. Moreover, for $d\geq4$,
$C_3$ is increasing when the radius of the circle is small relative to the
cut-off scale, \ie $\tilde{R}\lesssim\delta$, while it starts to decrease when
the radius is large. Hence, we find $\partial_{\tR}C_3<0$ for large $\tR$, as
required, but $\partial_{\tR}C_3$ becomes positive for $\tR\lesssim\delta$.
However, this problematic behaviour can be explained, as before, by the finite
resolution intrinsic to the RS2 model. Our results for the entanglement entropy
are not meaningful when $\tR\lesssim\delta$ because the model cannot
effectively distinguish the degrees of freedom inside and outside of the
cylinder. Note, however, that $d=3$ is a special case with
$\partial_{\tR}C_3>0$ for all values of $\tR$. Clearly, this case requires
further explanation, which we reserve for the following section.

\subsection{Results for $d=2$ and 3}

Both the slab geometry for $d=2$ and the cylindrical geometry for $d=3$ are
special cases. In particular, both cases appear to be problematic from the
point of view of the analysis in this section. We found above that
$\del_{\tR}C_3>0$ for all radii in $d=3$ and below we will show that
$\del_{\tl}C_2>0$ for all separations in $d=2$. Another distinctive feature of
these two cases is that the calculations can be done completely analytically,
as they are both special cases of the analysis given in appendix \ref{sphere}.

Hence let us present the analytic results. For $d=2$, the entanglement entropy
for the slab geometry becomes
 \ba
 S_\mt{EE}&=&\frac{8}{\pi}
 \ct\,\log\left(\frac{\tl}{2\delta}+\sqrt{1+\frac{\tl^2}{4\delta^2}}
 \,\right)
 \labell{seed2}\\
 &\simeq&
 \frac{8}{\pi} C_T
 \[\log\left({\tl}/{\delta}\right)+\frac{\delta^2}{\tl^2}+\cdots\]
 \quad{\rm for}\
 \tl\gg\delta \, , \nonumber
%
 \ea
where $C_T$ is the central charge given by eq.~(\ref{effectc}). Given this
result for $S_\mt{EE}$, we find
 \bea
 C_2&=&\frac{8\ct}{\pi}
 \,\frac{\tl}{\sqrt{\tl^2+4\delta^2}}\qquad\ \,\simeq\frac{8\ct}{\pi}
 \[1-\frac{2\delta^2}{\tl^2}+\cdots\] \,,
 \labell{c2222}\\
 \partial_{\tl\,}C_2&=&\frac{32\ct}{\pi}
 \,\frac{\delta^2}{\(\tl^2+4\delta^2\)^{3/2}}
 \ \ \simeq \frac{32\ct}{\pi}
 \,\frac{\delta^2}{\tl^3}+\cdots\, ,
 \labell{c2223}
 \eea
where the approximate expressions apply for $\tl\gg\delta$. Similarly, we
obtain a simple expression for entanglement entropy for cylindrical geometry in
$d=3$
 \bea
S_\mt{EE}&=&8\ct\,\left(\sqrt{\frac{\tilde{R}^2}{\delta^2}+1}- 1\right)
 \labell{seed3}\\
&\simeq&8\ct\,\[\frac{\tR}{\delta}-1+\frac{\delta}{2\tR}+\cdots\]
 \quad{\rm for}\
 \tR\gg\delta\,.\nonumber
 \eea
We use this expression for $S_\mt{EE}$ to calculate
 \bea
 C_3&=&8\ct\(1-
 \,\frac{\delta}{\sqrt{\tR^2+\delta^2}}\)
 \ \ \simeq8\ct\[1-\frac{\delta}{\tR}+\cdots\]\,,
 \labell{c3333}\\
 \partial_{\tR}C_3&=&8\ct
 \,\frac{\tR\, \delta\ \ }{\(\tR^2+\delta^2\)^{3/2}}
 \qquad\ \ \,\simeq 8\ct\,\frac{\delta}{\tR^2}+\cdots\,.
 \labell{c3334}
 \eea
We have again also presented the leading terms in an expansion for
$\tR\gg\delta$.

Since the expressions in eqs.~\reef{c2223} and \reef{c3334} are both positive,
it is evident that the inequalities in eqs.~\reef{ineq2}  and \reef{ineq3} are
never satisfied in these two cases. Further, as noted before,  it is clear the
finite resolution $\delta$ will not resolve this discrepancy since these
violations occur for arbitrarily large regions. A common feature of both of
these cases is that the gravity theory on the brane is somewhat unusual, \ie
for $d=2$ and 3, there will be no propagating graviton modes on the brane.
While this feature may make these theories seem somewhat pathological, we do
not believe that the failure of the inequalities is tied to this peculiar
property. In particular note that, with the slab geometry, we still found that
eq.~\reef{ineq2} is satisfied for $d=3$.

Instead, examining the large size expansions in
eqs.~(\ref{seed2}--\ref{c3334}), we find that in these two special cases, the
inequalities are probing contributions to the entanglement entropy that contain
positive powers of the cut-off (in the long-distance expansion). That is,
eq.~\reef{c2223} is controlled by the $\delta^2/\tl^2$ term in eq.~\reef{seed2}
for large $\tl$, while the $\delta/\tR$ term in eq.~\reef{seed3} dominates the
result in eq.~\reef{c3334} at large $\tR$. This contrasts to the cases where
eqs.~\reef{ineq2} and \reef{ineq3} were satisfied. As shown in eq.~\reef{c22}
for the slab geometry, we found the leading contribution to $C_2$ was
independent of $\delta$. For the cylindrical entangling surface, eq.~\reef{c32}
shows that the leading contribution to $C_3$ is controlled by $\kappa_3$,
which is proportional to $1/\delta^{d-4}$ for $d>4$ and to $\log\delta$ for
$d=4$. Further, we might note that such contributions with positive powers of
$\delta$ would be dropped in standard AdS/CFT calculations because they vanish
in the limit $\delta\to0$. Let us also observe that similar terms are also
becoming important where the previous calculations fail to satisfy the desired
inequalities, \ie when $\tl,\tR\lesssim\delta$.

Hence the calculations of $\partial_{\tl}C_2$ for $d=2$ and $\partial_{\tR}C_3$
for $d=3$ are scrutinizing the RS2 model in an essentially different way from
the previous calculations. In particular, the problems with eqs.~\reef{ineq2}
and \reef{ineq3} indicate that we are probing the RS2 model beyond its proper
regime of validity. We expect that the culpable feature in our framework
responsible for this bad behaviour is the superficial treatment of the cut-off
$\delta$ as a discrete surface in the AdS bulk. For example, in a stringy
construction \cite{herman,string}, the AdS space would extend smoothly into
some complex UV geometry. Of course, understanding the dual description of such
a construction would also be more difficult. In particular, an interesting
question would be finding the appropriate definition of the holographic
entanglement entropy to replace eq.~\reef{define}. Given the conjecture of
\cite{new1}, it seems that one should simply consider applying the BH formula
\reef{prop0} to some surface in the extended geometry. However, it remains to
find some principle that would select the appropriate surface in the UV
geometry. Given this reasoning, another perspective on our problems with
eqs.~\reef{ineq2} and \reef{ineq3} would to say that the standard holographic
prescription \reef{define} for the entanglement entropy must be supplemented by
order $\delta$ corrections when calculating $S_\mt{EE}$ in the RS2 braneworld
--- not a particularly surprising conclusion.


\section{Discussion} \labell{discuss}

In this paper, we used the Randall-Sundrum II braneworld as a framework to
study the conjecture \cite{new1} that in quantum gravity, the entanglement
entropy of a general region should be finite and the leading contribution is
given by the Bekenstein-Hawking area law \reef{prop0}. As this braneworld model
has a dual description in terms of gravity in an AdS bulk, we were able to
apply the usual prescription for holographic entanglement entropy to show that
this conjecture is realized in this model. The validity of this result required
that the curvatures of the brane geometry were small relative to the cut-off
scale, as in eq.~\reef{cond1}. Further, the geometry of the entangling surface,
\ie the boundary of the region for which $S_\mt{EE}$ is being calculated, must
also be sufficiently smooth as expressed in eq.~\reef{cond2}.

The entanglement entropy of general regions also shows interesting structure
beyond the area law term. In section \ref{region}, we extended our holographic
calculations to find the leading corrections to the BH term, which involve
integrals of background and extrinsic curvatures over the entangling surface.
One notable feature of the general result shown in eq.~\reef{entJM0} is that
the (dimensionful) coefficients of these correction terms in $S_\mt{EE}$ can be
expressed in terms of the gravitational couplings of the curvature-squared
coefficients in the induced gravity action. The latter action was derived in
appendix \ref{effact} and the general form of our results is given in
eq.~\reef{EGBind00}. It is natural to compare the Wald entropy \reef{wald} of
this gravity action with the entanglement entropy and we found
 \be
S_\mt{EE}=S_\mt{Wald}-\int_{\tS} d^{d-2}y \sqrt{  \hh} \[ \kappa_1\,K^iK_i+
 4\,\kappa_2\(K^i_{ab}K_i{}^{ab}-\frac{1}{d-2}K^iK_i\)\]+\cdots\,.
 \labell{boat}
 \ee
That is, $S_\mt{Wald}$ and $S_\mt{EE}$ match except that the extrinsic
curvature terms appearing in the entanglement entropy are absent in the Wald
entropy. However, since the extrinsic curvatures of a Killing horizon vanish,
this means that we will find $S_\mt{EE} = S_\mt{Wald}$ if the entanglement
entropy is evaluated on such a horizon, \eg of a stationary black hole. Of
course, this conclusion reinforces the results of \cite{rob06,andy} that
horizon entropy can be interpreted as entanglement entropy in the RS2 model.
Further, our result is perhaps natural given that the `off-shell' approach
\cite{important} to evaluating horizon entropy is constructed to take the form
of an entanglement entropy calculation and further when this approach is
applied in a higher curvature gravity theory, it reproduces precisely the Wald
entropy \cite{cthem}. Given that the extrinsic curvature terms in $S_\mt{EE}$
also appear multiplied by the gravitational couplings, it would be interesting
to construct an analogous `derivation' which also produces these terms for a
general horizon or a generic entangling surface.

As an indication of the robustness of these results, we compare
eq.~\reef{entJM0} with a perturbative calculation of the holographic entropy
functional for a general curvature-squared gravity action in the bulk
\cite{gbee}. Following the reasoning of \cite{new1}, this entropy functional
should represent the leading contribution to the entanglement entropy for
general regions in the AdS spacetime. Hence it is interesting to compare the
result emerging from the two different calculations for consistency. Their
analysis begins with a general curvature-squared action for a five-dimensional
gravity action, which for convenience we write as
 \beq
I = \frac{1}{16\pi G_5} \int \mathrm{d}^5x \, \sqrt{-g}\, \left[ \frac{12}{L^2}
+ R + L^2\left(\la_1\, C_{ijkl} C^{ijkl} +\la_2\,R_{ij}R^{ij}+\la_3\,R^2
\right)\right]\ .
 \labell{actorg}
 \eeq
The (dimensionless) couplings of the curvature-squared terms were assumed to be
small, \ie $\lambda_{1,2,3}\ll1$, and the calculations were only carried to out
to linear order in these couplings. Note that the action above contains a
negative cosmological constant term and so the vacuum solution is an AdS$_5$
spacetime. Considering the AdS/CFT correspondence in this context, the
objective in \cite{gbee} was to determine the appropriate prescription for
holographic entanglement entropy. By demanding that this prescription produce
the correct universal contribution to the entanglement entropy in the dual
four-dimensional CFT, as appears in eq.~\reef{donkey}, the following entropy
functional was constructed
 \beqa
S_\mt{EE} &=& \frac{\A(\sigma)}{4G_5}+\frac{L^2}{4G_5} \int_\sigma
d^{3}x\sqrt{h}\,\left[2\la_1\(h^{ac}h^{bd}\,C_{abcd}-K^i_{ab}K_i{}^{ab}\)
\right.\nonumber\\
&&\qquad\qquad\qquad\qquad\qquad\qquad \left. +\ \la_2\, R^{ij}{g}^{\perp}_{ij}
+2\la_3\, R + \alpha\, K^i K_i \right]\,,
 \labell{form55}
 \eeqa
where $\sigma$ denotes the extremal surface in the AdS bulk. Now comparing this
result with eq.~\reef{entJM0} with $d=5$, we find agreement for the leading
area term, of course, and further the terms involving the background curvatures
match the Wald entropy in both expressions. A more interesting observation is
that the coefficient of the $K^i_{ab}K_i{}^{ab}$ term precisely matches in both
expressions, \ie this coefficient is the same as that of the Weyl curvature
term but with the opposite sign. Unfortunately, no comparison can be made for
the $K^i K_i$ term because the coefficient $\alpha$ above remains undetermined
in eq.~\reef{form55}. This ambiguity arises because the calculations yielding
eq.~\reef{form55} were only linear in the higher curvature couplings, whereas
fixing $\alpha$ would require a higher order calculation because the leading
order equations extremizing the surface set $K^i=0$.\footnote{The suggestion
was made in \cite{gbee} to set $\alpha=2\lambda_1$ in order to simplify the
equations determining the extremal bulk surfaces. Of course, this choice would
disagree with the results in eq.~\reef{entJM0}.} However, the fact that the two
independent calculations agree on the coefficient of the $K^i_{ab} K_i{}^{ab}$
term seems to hint at the universal structure of the extrinsic curvature
contributions in $S_\mt{EE}$. It is also revealing that there are no additional
contributions to $S_\mt{EE}$ of this form for the action \reef{actorg} where
the couplings $\lambda_2$ and $\lambda_3$ are completely independent, whereas
with $d=5$, we have $\lambda_3=-\frac{5}{16}\lambda_2$ in eq.~\reef{EGBind00}.

Given eq.~\reef{boat}, it is interesting to examine the sign of the extrinsic
curvature corrections to $S_\mt{EE}$. For simplicity, let us assume that we are
considering the entangling surface on a fixed time slice in a stationary
background, \ie the time-like normal will not contribute to the extrinsic
curvatures. In this case, both of the geometric expressions in eq.~\reef{boat}
are positive (or vanishing).\footnote{If we denote the eigenvalues of
$K^i_{ab}$ for the space-like normal as $k_\al$, then $K^iK_i=\(\sum_\al
k_\al\)^2$ and $K^i_{ab}K_i{}^{ab}-\frac{1}{d-2}K^iK_i = \sum_\al k_\al^2
-\frac{1}{d-2}\(\sum_\al k_\al\)^2$. The latter can be shown to be positive or
zero using Lagrange's identity.} Hence the sign of the extrinsic curvature term
depends on the sign of the gravitational couplings, $\kappa_{1}$ and
$\kappa_2$. In particular, $S_\mt{EE}\le S_\mt{Wald}$ for
$\kappa_{1},\kappa_2\ge 0$. Hence this inequality is satisfied for the RS2
model constructed with Einstein gravity in the AdS bulk --- see
eq.~\reef{ekappa1}. However, the couplings for the RS2 model with GB gravity in
the bulk are given in eqs.~\reef{kappa1} and \reef{kappa2} and in this case, it
is clear $\kappa_2$ will be negative when the GB coupling $\lambda$ is
negative. A closer examination also shows that $\kappa_1$ will become negative
in $d\ge5$ if $\lambda$ becomes sufficiently positive. Hence for these models,
the extrinsic curvature corrections in eq.~\reef{boat} do not have a definite
sign. Of course, in dynamical circumstances, \eg in a cosmological setting or
for an expanding black hole, the time-like normal will also generically
contribute nonvanishing $K^t_{ab}$ and in such a situation, the geometric
expressions in eq.~\reef{boat} are no longer guaranteed to be positive. Hence
it does not possible to make a general statement about the sign of the
extrinsic curvature corrections and hence about the relative magnitude of
$S_\mt{EE}$ and $S_\mt{Wald}$.

It may seem desirable to establish an inequality of the form $S_\mt{EE}\le
S_\mt{Wald}$ as this would be inline with the intuitive statement that `black
holes are the most entropic objects' in the corresponding gravity theory, as
might arise in discussions of the Bekenstein bound \cite{beke0} or holographic
bounds \cite{bousso} on the entropy. Hence although the conjecture of
\cite{new1} suggests that in theories of quantum gravity, $S_\mt{EE}$ is finite
and closely related to the Bekenstein-Hawking entropy \reef{prop0}, the
previous discussion seems to indicate that entanglement entropy alone is not
the correct quantity in which to frame such discussions. In particular, in
examining entropy bounds, it seems crucial to relate the appropriate entropy
density to the stress-energy tensor \cite{flan}, which would not be achieved
by, \eg quantum correlations in the vacuum. Hence it seems a more refined
measure of the entropy is required for such discussions \cite{beke1}.

As an aside, let us add that \cite{furx,furx2} suggested that extremal surfaces
should play an important role in combining entanglement entropy and quantum
gravity. That is, the leading contribution to entanglement entropy should be
given by the BH formula \reef{prop0} but only when the entangling surface is an
extremal surface. This contrasts with the present perspective \cite{new1} where
extremal surfaces do not seem to play a special role. Certainly, our
calculations in the RS2 model establish $S_\mt{EE}={\cal A}/(4G_d)+\cdots$ for
arbitrary surfaces, not only event horizons. Further, while $K^i=0$ for an
extremal surface, this does not eliminate all of the extrinsic curvature
corrections in eq.~\reef{boat}.

As a final note, we remind the reader of the various limitations appearing in
our calculations. First of all, our results in eqs.~\reef{EGBind00} and
\reef{entJM0} rely on the geometries of both the background and the entangling
surface being weakly curved, as described by the constraints in
eqs.~\reef{cond1} and \reef{cond2}. Further, the calculations in section
\ref{test} for $d=2$ and 3 revealed new limitations, in that, contributions to
the entanglement entropy at $O(\delta/R)$ appear unreliable. It would appear
that this problem could be resolved by considering a stringy construction
\cite{herman,string} which emulates the RS2 model. In particular, such a
construction would give a better understanding of the geometric cut-off in the
AdS geometry. It would be interesting if this approach also gave some new
insights into the standard holographic prescription \reef{define} for
entanglement entropy. The discussion in section \ref{test} also showed that
there are basic limitations to assigning an entanglement entropy to spacetime
regions, which are generic rather than being specific to the RS2 model. In
particular, one expects that any theory of quantum gravity will only
distinguish different regions of spacetime with some finite resolution and so
one will not be able to meaningfully assign an entanglement entropy to
arbitrarily small regions (or regions defined by geometric features which are
arbitrarily small). We note that the assumptions of strong subadditivity,
Lorentz symmetry and causality lead one to conclude that if the entanglement
entropy of any arbitrary region in flat space is finite then it must be given
by precisely $S_\mt{EE}=c_0 {\cal A} +c_1$, where $c_0$ and $c_1$ are universal
constants \cite{Carea}. Hence the `failure' of the putative entanglement
entropy for arbitrarily small regions in section \ref{test} is actually an
essential ingredient to providing a nontrivial result \reef{entJM0} at large
scales.

\acknowledgments

We would like to thank Eugenio Bianchi, Raphael Bousso, Horacio Casini, Ben
Freivogel, Dmitry Fursaev, Matt Headrick, Simeon Hellerman, Janet Hung, Markus Luty, Masimo
Porrati, Sergey Solodukhin and Herman Verlinde for useful conversations.
Research at Perimeter Institute is supported by the Government of Canada
through Industry Canada and by the Province of Ontario through the Ministry of
Research \& Innovation. RCM also acknowledges support from an NSERC Discovery
grant and funding from the Canadian Institute for Advanced Research.

\appendix


\section{Induced Gravity Action}
\label{effact}

In this appendix, we use the Fefferman-Graham expansion given in
eqs.~\reef{expandfg} and \reef{expand} to explicitly evaluate the first few
contributions in the derivative expansion of the induced gravity action
\reef{bulkact2} on the brane. In the following, we consider a bulk theory with
higher curvature interactions, namely Gauss-Bonnet (GB) gravity \cite{lovel}.
One should regard this theory as a toy model which may provide some insights
into more general holographic CFT's. In particular, having a curvature-squared
term in the bulk results in the boundary theory having two independent central
charges \cite{highc}. In part, this feature motivated several recent
holographic studies of GB gravity, \eg \cite{cc,EtasGB}. Of course, the results
for Einstein gravity \reef{EHact} are easily obtained from the following by
taking the limit where the higher curvature coupling vanishes.

The GB gravity action in the bulk takes the form\footnote{As in the main text,
calligraphic $\R$ and $\K$ are used to denote bulk curvature and the second
fundamental form of the brane, respectively. Recall that there are two copies
of the AdS geometry and so implicitly, we assume that bulk integral runs over
both copies and surface integral is carried over both sides of the brane.}
 \be
I_{bulk}^\mt{GB}=\frac {1}{16\pi G_{d+1}}\int d^{d+1}\!x\, \sqrt{-G}
\left[\frac{d(d-1)}{L^2}+\R+\frac{{L}^2\,\lambda}{(d-2)(d-3)}\,\chi_4
\right]+I^\mt{GB}_{surf}\,.
 \labell{GBactx}
 \ee
where $\chi_4$ is proportional to the four-dimensional Euler density,
 \beq
\chi_4=\R_{\mu\nu\rho\sigma}\R^{\mu\nu\rho\sigma}-4\,
\R_{\mu\nu}\R^{\rho\sigma}+\R^2\,. \labell{euler4x}
 \eeq
This curvature-squared interaction in the bulk requires higher curvature
contributions in the surface action \cite{surf}, which appears along with the
standard Gibbons-Hawking-York term for the Einstein-Hilbert action,
 \bea
I^\mt{GB}_{surf}&=&{1\over 16\pi G_{d+1}}\int d^{d}x\, \sqrt{-\tg} \,\[ 2\,\K +
\frac{{L}^2\,\lambda}{(d-2)(d-3)}\( 4 R\,\K-8 R_{ij}\K^{ij} \vphantom{{4\over
3}} \right.\right.
\labell{GBsurf}\\
&& \qquad\qquad\qquad\qquad\qquad\qquad\qquad\qquad \left.\left. -{4\over
3}\K^3 +4\K\K_{ij}\K^{ij}-{8\over 3}\K_{ij}\K^{jk}\K^{i}{}_k\)\]\, ,
 \nonumber
 \eea
where $\tg_{ij}$ corresponds to the induced metric on the brane.

While $L$ sets the scale of the cosmological constant in eq.~\reef{GBactx}, one
easily finds that the AdS curvature scale is actually given by
  \be
\delta^2=\tilde L^2={L^2\over f_\infty}\qquad{\rm where}\ \
f_\infty={1-\sqrt{1-4\lambda}\over 2\lambda}~.
 \labell{oxcart}
 \ee
Here we are using the relation $\delta=\tL$ which holds for the RS2 model, as
discussed in section \ref{ransun}. Note that we chosen $\fin$ such that with
$\lambda\to 0$, $f_\infty=1$ and so we recover $\tL=L$ in this limit.
Implicitly, $f_\infty$ is determined as the root of a quadratic equation and we
are discarding the other root since with this choice, the graviton would be a
ghost and hence the dual CFT would not be unitary \cite{GBghost,old1}. Further,
constraints on the holographic construction limit the GB coupling to lie in the
following range, \eg \cite{cc}
 \be
-\frac{(3d+2)(d-2)}{4(d+2)^2}\le\lambda\le
\frac{(d-2)(d-3)(d^2-d+6)}{4(d^2-3d+6)^2}
 \labell{constra}
 \ee
for $d\ge4$. As noted above, one interesting feature of GB gravity
\reef{GBactx} is that the dual boundary theory will have two distinct central
charges. Following \cite{cthem,cc}, we define these charges as:\footnote{For
convenience, our normalizations of $\ct$ and $\ads$ are slightly different here
than originally appears in \eg \cite{cthem,cc}.}
 \bea
\ct&=& \frac{\pi}{8}\, \frac{\delta^{d-1}}{ G_{d+1}} \left[ 1 - 2 \lambda \fin
\right]\,,
 \labell{effectc}\\
\ads &=& \frac{\pi}{8}\, \frac{\delta^{d-1}}{ G_{d+1}} \left[ 1 -
2\frac{d-1}{d-3} \lambda \fin \right]\,. \labell{effecta}
 \eea
The first charge $\ct$ controls the leading singularity of the two-point
function of the stress tensor. The second central charge $\ads$ can be
determined by calculating the entanglement entropy across a spherical
entangling surface \cite{cthem}. In even dimensions, $\ads$ is also
proportional to the central charge appearing in the A-type trace anomaly
\cite{cthem}. Note that in the limit $\la\to0$, $\ct=\ads$.

For GB gravity as presented in eq.~\reef{GBactx}, the two unknown coefficients
$k_1$ and $k_2$ in eq.~\reef{metricexpand2} are given by \cite{gbee}
 \bea
 k_1&=&{3\over 4(d-1)(d-2)(d-3)(d-4)} ~ {\lambda f_\infty\over (1-2\lambda f_\infty)}\,~,
 \nonumber\\
k_2&=&-\frac43\, (d-1)\ k_1~.
 \labell{k-coeff}
 \eea
Now the equations of motion for the metric in the bulk are given by
 \bea
 &&\R_{\mu\nu}-{G_{\mu\nu}\over 2}\bigg(\R+{d(d-1)\over L^2}+{L^2\lambda\over (d-2)(d-3)}
 \chi_4\bigg)
  \labell{here}\\
 &&\qquad\qquad
 +\frac{\,2 L^2\,\lambda}{(d-2)(d-3)}\big( \R_{\mu\sigma\rho\tau}\R_\nu{}^{\sigma\rho\tau}
 -2\R_{\mu\rho}\R_\nu{}^\rho-2\R_{\mu\rho\nu\sigma}\R^{\rho\sigma}+\R\R_{\mu\nu} \big)=0\, .
 \nonumber
 \eea
Taking trace of these equations then yields
 \be
 {L^2\lambda\over (d-2)(d-1)}\,\chi_4=-\R-{d(d+1)\over  L^2} \, .
 \labell{track}
 \ee
Hence, the on-shell bulk action  can be written as follows
 \be
2I_{bulk}^\mt{GB}=-\frac {1}{4\pi G_{d+1}(d-3)}\int d^{d+1}\!x\, \sqrt{-G}
\Big[\frac{2d(d-1)}{L^2}+\R \Big] + 2I^\mt{GB}_{surf}\,,
 \labell{GBonshell}
 \ee
where $I^\mt{GB}_{surf}$ is given in eq.~\reef{GBsurf}. We have included an
extra factor of two above, as in eq.~\reef{act1}, since we are assuming that
the integrals above run over one copy of the AdS space.

The outward-pointing unit normal at the cut-off surface, $\rho=1$, is given by
$n_{\mu}=-\sqrt{G_{\rho\rho}} \, \delta^\rho_\mu$. Now one can readily evaluate
derivative expansion of the second fundamental form at this surface
 \be
\mathcal{K}_{ij}=\nabla_i n_j |_{\rho=1}=-{\rho\over\delta} {\del
G_{ij}\over\del\rho}\Big|_{\rho=1}={1\over\delta} \sum_{n=0}^{\infty}
(1-n)\overset{\scriptscriptstyle{(n)}}{g}_{ij} ={1\over\delta}
\big(\tg_{ij}-\sum_{n=1}^{\infty}n\gn_{ij}\big)~,
 \labell{Kbrane}
 \ee
where we are using $\tL=\delta$. Recall that eq.~\reef{bond} gives the induced
metric $\tg_{ij}$ on the brane in terms of the FG expansion coefficients
\reef{expand}.

Now, the general expansion of the curvature scalar requires rather tedious
computations. However, we employ a shortcut since we will only carry the
derivative expansion to fourth order. In this case, we need only
$\overset{\scriptscriptstyle{(1)}}{g}\!_{ij}$ and
$\overset{\scriptscriptstyle{(2)}}{g}\!_{ij}$ in the FG expansion
\reef{expand}. The main observation for our shortcut is to exploit Einstein
gravity in order to argue that for any gravity theory in the bulk only terms
proportional to $k_1$ and $k_2$ in eq. \reef{metricexpand2} contribute
nontrivially at fourth order in the derivative expansion of the curvature
scalar while the second order term in such expansion vanishes independently of
the details of the bulk gravity theory.

Indeed, in the case of Einstein gravity (for which $\delta=\tL=L$), the Ricci
scalar is constant by the equations of motion, \ie eq.~\reef{track} yields
$\R=-d(d+1)/\delta^2$ (with $\la=0$). Therefore in the derivative expansion,
coefficients of all higher order corrections vanish. Furthermore, we observe
that $k_1=k_2=0$ from eq.~\reef{k-coeff} with $\lambda=0$. Hence we may deduce
that in the absence of $k_1$ and $k_2$, the contributions that originate from
$\overset{\scriptscriptstyle{(1)}}{g}\!_{ij}$ and
$\overset{\scriptscriptstyle{(2)}}{g}\!_{ij}$ cancel each other. Therefore with
a general theory for bulk gravity, only Weyl-squared terms in
eqs.~\reef{metricexpand} and \reef{metricexpand2} can contribute in a
nontrivial way at the fourth order in the derivative expansion of the curvature
scalar, whereas second order must vanish identically. Now since the
Weyl-squared terms already possess four derivatives, it is enough to perform
linear analysis to find the desired contributions in the expansion of $\R$.
That is, first we rewrite eq.~\reef{expandfg} as
 \be
ds^2= G_{\mu\nu}\,dx^\mu dx^\nu=\frac{\delta^2}{4}\frac{d\rho^2}{\rho^2} +
\frac{1}{\rho}\,\gz_{i j}(x)\,dx^i dx^j+\delta G_{ij}(x,\rho)\,dx^i dx^j\, ,
 \labell{gong}
 \ee
where, in principle, one has
 \be
\delta G_{ij}(x,\rho)=\go_{ij}(x)+\gs_{ij}(x)\rho+\cdots=\sum_{n=1}^\infty
\gn(x)\,\rho^{n-1}~. \labell{show}
 \ee
Then we can evaluate linear correction to $\R$ associated with $\delta G_{ij}$,
however, for the present purposes, we do not use the entire expression
\reef{show} but rather we keep only contributions of the Weyl-squared terms
appearing in $\overset{\scriptscriptstyle{(2)}}{g}\!_{ij}$.

The first variation of the curvature scalar under $G_{\mu\nu}\to
G_{\mu\nu}+\delta G_{\mu\nu}$ is given by
 \be
 \delta\R=-\R^{\mu\nu}\delta G_{\mu\nu}+\nabla^\mu(\nabla^\nu \delta G_{\mu\nu}
 -  \nabla_\mu \delta G^\nu{}_{\nu})~,
 \labell{deltaR}
 \ee
where covariant derivative $\nabla_{\mu}$ is compatible with unperturbed metric
$G_{\mu\nu}$ which is also used to raise and lower the indices in the above
expression. In our case, the unperturbed Ricci tensor is given by
 \be
 \R^{\rho\rho}=-{4d\over\delta^4}\rho^2 ~,\quad
 \R^{ij}=\rho\(\rho\,R^{ij}[\gz]-{d\over \delta^2}\gz^{ij}\)~,
 \labell{taxi}
 \ee
where indices in parenthesis are raised and lowered with
$\overset{\scriptscriptstyle{(0)}}{g}\!_{ij}$.  Combining the above results
altogether, we find the following expansion for the curvature scalar to fourth
order in the derivative expansion:
 \be
\mathcal{R}=-{d(d+1)\over \delta^2}+4(d-3) (d\,k_1+k_2)\, \delta^2\rho^2\, C_{m
n k l}C^{m n k l} +\cdots\,.
 \ee
In particular, in the special case of GB gravity \reef{GBact}, it follows from
eq. \reef{k-coeff} that
 \be
\mathcal{R}=-{d(d+1)\over \delta^2}- {1\over (d-1)(d-2)} ~ {\lambda
f_\infty\over (1-2\lambda f_\infty)}\,~\delta^2\rho^2~ C_{m n k l}C^{m n k l}
+\cdots\,.
  \labell{Rexpand}
 \ee

Next we substitute eqs.~\reef{Kbrane} and \reef{Rexpand} into
eqs.~\reef{GBsurf} and \reef{GBonshell} and then integrate over the extra
dimension $\rho$ in eq.~\reef{GBonshell}. The final result takes the form
 \bea
&&2I^\mt{GB}_{bulk}={\delta\over 8\pi (d-2) G_{d+1}}\int d^{d}x\sqrt{-\tg}
\[ {2(d-1)(d-2)\over \delta^2}\big(1-{2\over 3}\lambda
f_\infty\big)+(1+2\lambda f_\infty)R \right.
 \labell{dinal}\\
&&\ \
\left. \quad\quad +{1-6\lambda f_\infty\over (d-2)(d-4)}\, \delta^2\bigg(R_{ij}R^{ij}
 -{d\over 4(d-1)}R^2\bigg)+{\lambda f_\infty\over (d-3)(d-4)}\,\delta^2\,C_{ijkl}
 C^{ijkl}+\mathcal{O}(\del^6)  \] ~.
 \nonumber
 \eea
Note that implicitly the above expression only contains the contribution from
the lower limit of the $\rho$ integration, \ie from $\rho=1$. Our result
coincides with the $\I n$ terms  in eq.~\reef{bulkact2} for $n=0$, 1 and 2. Up
to the Weyl-squared term, this boundary action is identical to that found in
\cite{yale} for GB gravity. However, the Weyl-squared term was absent in
\cite{yale} simply because the analysis there only considers conformally flat
boundaries. To get the full induced gravity action \reef{bulkact2} on the
brane, we need to add $I_{brane}$ to the above expression. In the absence of
any matter fields, the latter has the simple form
 \be
I_{brane}= -T_{d-1}\int d^{d}x\sqrt{-\tg}\,.
 \labell{brain}
 \ee
Now for simplicity, we tune the brane tension to be
 \be
T_{d-1}  ={d-1\over 4\pi G_{d+1}\delta }\bigg(1-{2\over 3}\lambda f_\infty
\bigg)~,
 \labell{tensx}
 \ee
so that it precisely cancels the cosmological constant contribution in
eq.~\reef{dinal}. Combining these expressions together, we finally obtain
 \be
I^\mt{GB}_{ind}=\int d^{d}x\sqrt{-\tg} \[ {R\over 16\pi G_{d}}
+\,\frac{\kappa_1}{2\pi}\bigg(R_{ij}R^{ij}-{d\over 4(d-1)}R^2\bigg)
+\frac{\kappa_2}{2\pi} \,C_{ijkl} C^{ijkl}+\mathcal{O}(\del^6)  \] ~.
  \labell{EGBind}
 \ee
where the effective $d$-dimensional Newton's constant is given by
 \be
\frac{1}{ G_d}={2\,\delta\over d-2}{1+2\lambda f_\infty\over G_{d+1}}
 =\frac{16}{\pi(d-2)}\,\frac{(d-2)\ct-(d-3)\ads}{\delta^{d-2}} \, ,
 \labell{eegb2x}
 \ee
and the couplings for the curvature-squared terms can be written as
 \bea
 \kappa_1&=&{\delta^3\over 4(d-2)^2(d-4)}\, {1-6\lambda f_\infty\over
G_{d+1}} ={2\over \pi(d-2)^2(d-4)}\, {(d-3)\ads-(d-4)\ct\over \delta^{d-4}}\,,
 \labell{kappa1}\\
 \kappa_2&=&{\delta^3\over  4(d-2)(d-3)(d-4)}\,{\lambda f_\infty\over
G_{d+1}} = {1\over  2\pi(d-2)(d-4)}\,{\ct-\ads\over \delta^{d-4}}\,.
  \labell{kappa2}
\eea

Now setting $\la=0$ above, we recover the induced action for Einstein gravity
\reef{EHact} in the bulk
 \be
I^\mt{E}_{ind}=\int d^{d}x\sqrt{-\tg} \[ {R\over 16\pi G_{d}}
+\,\frac{\kappa_1}{2\pi} \bigg(R_{ij}R^{ij}-{d\over 4(d-1)}R^2\bigg)
+\mathcal{O}(\del^6)
\] ~,
 \labell{EHind}
 \ee
where the induced couplings can be written as
 \bea
\frac{1}{G_d}&=&{2\,\delta\over d-2}\,{1\over G_{d+1}}
 =\frac{16}{\pi(d-2)}\,\frac{\ct}{\delta^{d-2}}  \,,
 \labell{enewton}\\
\kappa_1&=&{\delta^3\over 4(d-2)^2(d-4)}\, {1\over  G_{d+1}} ={2\over
\pi(d-2)^2(d-4)}\, {\ct\over \delta^{d-4}}\,.
 \labell{ekappa1}
 \eea
Note that in this case, induced gravity action does not contain a term
proportional to the square of the Weyl tensor, \ie $\kappa_2=0$.

\section{Codimension-two Bulk Surfaces} \labell{geom}

In this appendix, we consider various curvatures associated with
codimension-two surface $\sigma$ in the bulk and evaluate their derivative
expansion up to second order. The formulae that we obtain here are useful in
the derivation of eq.~\reef{entJM}.

Recall that FG-like expansion of the induced metric on $\sigma$ was given in
eq.~\reef{inducemet}. Let us rewrite its components in the following way
 \be
 h_{\rho\rho}={\delta^2\over4\rho^2}+\delta h_{\rho\rho}\,,\qquad
 h_{ab}={\overset{\scriptscriptstyle{(0)}}{h}_{ab}\over \rho}+\delta h_{ab}\, .
 \labell{fire}
 \ee
Here, we are again using $\tilde L=\delta$, as is appropriate for calculations
in the RS2 model, and further we have defined
 \be
\delta h_{\rho\rho}={\delta^2\over4}\sum_{n=1}^\infty\hn_{\rho\rho}\,\rho^{n-2}
~,\qquad \delta h_{ab}=\sum_{n=1}^\infty\hn_{ab}\,\rho^{n-1}~.
 \labell{truck}
 \ee
As in eq.~\reef{taxi}, the Ricci tensor of the leading order metric
$\overset{\scriptscriptstyle{(0)}}{h}\!_{\al\beta}$ is given
by\footnote{Indices of Ricci tensor $R^{ij}[\hz]$ are raised and lowered with
$\overset{\scriptscriptstyle{(0)}}{h}\!_{ij}$.}
 \be
\R^{\rho\rho}=-{4(d-2)\over\delta^4}\rho^2 ~,\quad
\R^{ij}=\rho\big(\rho\,R^{ij}[\hz]-{(d-2)\over \delta^2}\hz^{ij}\big)~,
 \ee
Now applying eq.~\reef{deltaR} for the full induced metric \reef{fire} yields
 \bea
\mathcal{R}&=&-{(d-1)(d-2)\over \delta^2}+\rho\bigg(R_\Sigma+{(d-2)(d-3)\over
\delta^2}\ho_{\rho\rho}+{2(d-3)\over \delta^2}
\hz^{ab}\ho_{ab}\bigg)+\mathcal{O}(\del^4)
 \nonumber\\
&=&-{(d-1)(d-2)\over \delta^2}+\rho\bigg(R_\Sigma-{d-3\over d-2}\left[
2\hz^{ab}R_{ab}-{d-2\over d-1}\,R + K^iK_i \right] \bigg)+\mathcal{O}(\del^4)\, ,
\non
 \labell{Rsigma0}
 \eea
where we have explicitly substituted for
$\overset{\scriptscriptstyle{(1)}}{h}\!_{\al\beta}$ using
eqs.~\reef{metricexpand} and \reef{indmetric} in the second line. Here
$R_\Sigma$ denotes intrinsic curvature scalar for $\Sigma$. However, note that
to the order that we are working the latter is indistinguishable from the
intrinsic Ricci scalar evaluated for $\tS$, the entangling surface on the
brane, \ie using eq.~\reef{okax}, $R_\Sigma=R_{\tS}+O(\del^4)$.

To evaluate the holographic entanglement entropy in section \ref{GBtangle}, it
is useful to apply further geometric identities to re-express the first order
term in eq.~\reef{Rsigma0}. In particular, we use the Gauss-Codazzi equation
 \be
[R_{\tS}]_{abcd}=R_{abcd} +K^i_{ac}K_{i\,bd} -K^i_{ad}K_{i\,bc}
 \labell{GCrel}
 \ee
along with
 \be
\hh^{ac}\hh^{bd}C_{abcd}=\hh^{ac}\hh^{bd}R_{abcd}-{2(d-3)\over
d-2}\hh^{bd}R_{bd}+{d-3\over d-1}R\,,
 \labell{weyl}
 \ee
where $C_{ijkl}$ denotes the Weyl tensor evaluated with the brane metric.
Combined these identities allow us to re-express eq.~\reef{Rsigma0} as
 \be
\mathcal{R} =-{(d-1)(d-2)\over \delta^2}+\rho\left(\hh^{ac}\hh^{bd} C_{abcd}
-K^i_{ab}K_i{}^{ab} +\frac{1}{d-2}K^iK_i \right) +\mathcal{O}(\del^4) \, .
 \labell{Rsigma1}
 \ee

For the present purposes, the entangling surface $\tS$ is the boundary of the
extremal surface $\sigma$ and so we now turn to evaluate the second fundamental
form with the above asymptotic expansion. The outward normal vector of $\tS$
imbedded into $\sigma$ is $n_{\al}=-\sqrt{h_{\rho\rho}} \, \delta^\rho_\al$.
Hence, extrinsic curvature tensor takes the following form
 \be
\mathcal{K}_{ab}=\nabla_a n_b=-{1\over 2\sqrt{h_{\rho\rho}}} {\del h_{a
b}\over\del\rho}\Big|_{\rho=1}={\overset{\scriptscriptstyle{(0)}}{h}_{ab}\over  \delta} \big(1-{1\over
2}\ho_{\rho\rho}\big) +\mathcal{O}(\del^4)~,
 \labell{frog}
 \ee
whereas its trace is given by
 \bea
\mathcal{K}&=&{d-2\over  \delta} \big(1-{1\over
2}\ho_{\rho\rho}\big)-{1\over\delta} \hz^{ab}\ho_{ab}+\mathcal{O}(\del^4)
 \nonumber\\
&=&{d-2\over  \delta}-\frac{\delta}{2(d-2)}\left(
2R^{ij}\,{\tg}^{\perp}_{ij}-{d\over d-1}\,R- K^iK_i \right)
+\mathcal{O}(\del^4) \,.
 \labell{Ksigma}
 \eea
In the second line, we have explicitly substituted for
$\overset{\scriptscriptstyle{(1)}}{h}\!_{\al\beta}$ using
eqs.~\reef{metricexpand} and \reef{indmetric}. We have also simplified the
resulting expression using $R=R^{ab}\, \hh_{ab}+R^{ij}\,{\tg}^{\perp}_{ij}$.

\section{Spherical Entangling Surfaces} \labell{sphere}

In this appendix, we compare our perturbative results for the entanglement
entropy in section \ref{region} with those for a simple case where the entire
holographic result can be calculated analytically, namely a spherical
entangling surface in flat space. For this purpose, we consider the case where
the bulk is pure AdS space and the brane geometry is flat. In this situation,
we have $g_{ij}(x,\rho)=\eta_{ij}$ in eq.~\reef{expandfg} and the full metric
coincides with the standard Poincar\'e patch metric upon substituting
$z^2=\delta^2\rho$. Further, choosing the entangling surface $\Sigma$ in the
AdS boundary to be a $(d-2)$-dimensional sphere of radius $R$, then the
extremal surface $\sigma$ is given by \cite{rt1}
 \be
\delta^2\,\rho+ r^2 = R^2 = \tilde{R}^2+\delta^2 \, ,
 \labell{circle}
 \ee
where $r$ is the radial coordinate in the boundary geometry. Here we have also
introduced $\tilde R$, which corresponds to the radius of the spherical
entangling surface $\tS$ on the brane, \ie at $\rho=1$. In fact, the derivation
of \cite{casini9} shows that this same surface will be the appropriate extremal
surface, independently of the bulk gravity theory. As it will prove useful
below, let us write the induced metric on $\sigma$
 \be
ds^2=\frac{\delta^2}4\,\frac{d\rho^2}{\rho^2}\left(1+\frac{\delta^2}{r^2}
\,\rho \right) + \frac{r^2}{\rho}\,d\Omega^2_{d-2}\,.
 \labell{inner}
 \ee

Now in the case of Einstein gravity in the bulk, the holographic prescription
\reef{define} yields the following \cite{rt1}
 \bea
S_\mt{EE}&=&2\,\frac{\A(\sigma)}{4G_{d+1}}=\frac{\delta^{d-1}}{2G_{d+1}}
\,\Omega_{d-2}\int_{\delta\over\sqrt{\delta^2+\tilde{R}^2}}^1 dy
{(1-y^2)^{d-3\over 2}\over y^{d-1}}
 \labell{apple}\\
&=& \frac{\delta^{d-1}\Omega_{d-2}}{2G_{d+1}}
\[ { (1+\tilde{R}^2/\delta^2)^{d-2\over 2}\over d-2}
\,_2F_1\({2-d\over 2},{3-d\over 2},{4-d\over 2}, {1\over 1+\tilde{R}^2/\delta^2}\)
+{\Gamma\big({2-d\over 2}\big)\Gamma\big({d-1\over 2}\big) \over 2\sqrt\pi}
\] \, ,
 \nonumber
 \eea
where we have again introduced a factor of two above to account for the two
copies of the AdS geometry and $\Omega_{d-2}$ is the surface area of a
$(d-2)$-dimensional sphere of unit radius, \ie $\Omega_{d-2} =
2\pi^{(d-1)/2}/\Gamma\({d-1\over 2}\)$. Now to satisfy the constraint
\reef{cond2}, we consider a large sphere with $\tilde{R}\gg\delta$. In this
case, we may expand the result in eq.~\reef{apple} to find
 \be
S_\mt{EE}=\frac{{\cal A}(\tS)}{4G_d}\( 1 -{d-2\over 2(d-4)} \({\delta\over
\tilde R}\)^2+\cdots \)~,
 \labell{peach}
 \ee
where we substituted for the $d$-dimensional Newton's constant using
eq.~\reef{enewton} and we wrote ${\cal A}(\tS)=\Omega_{d-2}{\tilde R}^{d-2}$
for the area of the entangling surface. Hence we again find the leading term
takes precisely the form of the BH entropy \reef{prop0}. Further let us match
the first correction to that in eq.~\reef{EHent}. First, we calculate the
extrinsic curvatures of the sphere of radius $\tilde R$ as
 \be
K^{\hat t}_{ab}=0\qquad{\rm and}\qquad \quad K^{\hat r}_{ab}={\delta_{ab}\over \tilde R}~,
\labell{Sextcurv}
 \ee
where the first is associated with a time-like normal vector $n^{\hat
t}_i=\delta^t_i$ and the second with the radial normal $n^{\hat
r}_i=\delta^r_i$. Now using eq.~\reef{ekappa1}, we find there is a precise
agreement between the first corrections appearing in eqs.~\reef{EHent} and
\reef{peach}.

Let us now turn to the case of Gauss-Bonnet gravity \reef{GBact}. Now for the
holographic calculation of entanglement entropy, we extremize the new entropy
functional in eq.~\reef{eegb}. However, as noted above, for a spherical
entangling surface $\tS$ in the boundary theory, the extremal surface $\sigma$
in the bulk is again given by eq.~\reef{circle}. Hence we must examine the
geometry of this surface somewhat more closely to evaluate the desired
$S_\mt{JM}$. First of all, although it is not immediately evident from
eq.~\reef{inner}, $\sigma$ is a constant curvature surface with
 \be
 {\cal R}= -(d-1)(d-2)/\delta^2\,.
 \labell{RRR}
 \ee
Next, the extrinsic curvature of the boundary $\partial\sigma$ on the brane,
\ie $\rho=1$ is given by
 \be
{\cal K}_{ab} = - \frac{1}{2\sqrt{h_{\rho\rho}}}\,\left. \frac{\partial
h_{ab}}{\partial
\rho}\right|_{\rho=1}=\frac{\hh_{ab}}{\delta}\sqrt{1+\frac{\delta^2}{{\tilde
R}^2}}\,.
 \labell{KKKK}
 \ee
As shown in \cite{gbee}, combining these results yields
 \be
S_{JM}=\[1-2{d-1\over d-3}\lambda f_{\infty}\]{\A(\sigma)\over 2G_{d+1}}+{2
\lambda f_{\infty}\over d-3}\,{\delta \over
G_{d+1}}\sqrt{1+\frac{\delta^2}{{\tilde R}^2}} \ \Omega_{d-2} \tilde{R}^{d-2}~,
 \labell{banana}
 \ee
where the formula for $\A(\sigma)$ is the same as in the case of Einstein
gravity eq.\reef{apple}. As above, we expand this expression for
$\tilde{R}\gg\delta$ and the result may be written as
 \be
S_\mt{EE}=\frac{{\cal A}(\tS)}{4 G_d}\Big( 1 -{1-6\lambda f_\infty\over
1+2\lambda f_\infty}\,{d-2\over 2(d-4)}  \({\delta\over \tilde R}\)^2+\cdots
\Big)~,
 \labell{lemon}
 \ee
after substituting with eq.~\reef{eegb2x}. Now examining the previous result in
eq.~\reef{entJM}, we first note that the combination of extrinsic curvatures
appearing in the $\kappa_2$ term vanishes if we substitute with
eq.~\reef{Sextcurv}. However, using eqs.~\reef{kappa1} and \reef{Sextcurv}, we
find an exact agreement between the $\kappa_1$ term appearing in
eq.~\reef{entJM} and the first correction appearing above in eq.~\reef{lemon}.



\begin{thebibliography}{99}

\bibitem 
 {beks} J.~D.~Bekenstein,
``Black holes and the second law,''
  Lett.\ Nuovo Cim.\  {\bf 4}, 737 (1972);\\
 J.~D. Bekenstein, ``{Black holes and entropy},''
    Phys. Rev. D {\bf 7}, 2333 (1973);\\
 J.~D.~Bekenstein,
  ``Generalized second law of thermodynamics in black hole physics,''
  Phys.\ Rev.\ D {\bf 9}, 3292 (1974).

\bibitem 
 {area} S.~W.~Hawking,
  ``Black holes in general relativity,''
  Commun.\ Math.\ Phys.\  {\bf 25}, 152 (1972);\\
  S.~W.~Hawking and G.~F.~R.~Ellis,
  {\sl The Large scale structure of space-time},
  (Cambridge University Press, Cambridge, 1973).

\bibitem 
 {hawk} S.~W. Hawking, ``{Black hole explosions},'' {Nature} {\bf 248}, 30 (1974);\\
S.~W. Hawking, ``{Particle Creation by Black Holes},'' {Commun. Math.
  Phys.} {\bf 43}, 199 (1975).

\bibitem 
 {four} J.~M.~Bardeen, B.~Carter and S.~W.~Hawking,
  ``The Four laws of black hole mechanics,''
  Commun.\ Math.\ Phys.\  {\bf 31}, 161 (1973).


\bibitem 
 {DS} G.~W.~Gibbons and S.~W.~Hawking, ``Cosmological event
    horizons, thermodynamics, and particle creation,'' Phys. Rev. D {\bf 15},
    2738 (1977).

\bibitem 
 {ray} R.~Laflamme,
  ``Entropy Of A Rindler Wedge,''
  Phys.\ Lett.\ B {\bf 196}, 449 (1987).

\bibitem 
 {WaldEnt} R.~M.~Wald, ``Black hole entropy is the
    Noether charge,''
  Phys.\ Rev.\  D {\bf 48}, 3427 (1993)
  [arXiv:gr-qc/9307038];\\
T.~Jacobson, G.~Kang and R.~C.~Myers,
  ``On Black Hole Entropy,''
  Phys.\ Rev.\  D {\bf 49}, 6587 (1994)
  [arXiv:gr-qc/9312023];\\
V.~Iyer and R.~M.~Wald, ``Some properties of Noether charge and a proposal for
dynamical black hole entropy,''
  Phys.\ Rev.\  D {\bf 50}, 846 (1994)
  [arXiv:gr-qc/9403028].

\bibitem 
 {new1} E.~Bianchi and R.~C.~Myers,
  ``On the Architecture of Spacetime Geometry,''
  arXiv:1212.5183 [hep-th].

\bibitem 
 {luty} J.~H.~Cooperman and M.~A.~Luty,
  ``Renormalization of Entanglement Entropy and the Gravitational Effective Action,''
  arXiv:1302.1878 [hep-th].

\bibitem 
 {rt1} S.~Ryu and T.~Takayanagi,
  ``Holographic derivation of entanglement entropy from AdS/CFT,''
  Phys.\ Rev.\ Lett.\  {\bf 96}, 181602 (2006)
  [arXiv:hep-th/0603001];\\
 S.~Ryu and T.~Takayanagi,
  ``Aspects of holographic entanglement entropy,''
  JHEP {\bf 0608}, 045 (2006)
  [arXiv:hep-th/0605073];\\
 T.~Nishioka, S.~Ryu and T.~Takayanagi,
  ``Holographic Entanglement Entropy: An Overview,''
  J.\ Phys.\ A  {\bf 42}, 504008 (2009)
  [arXiv:0905.0932 [hep-th]].

\bibitem 
 {arealaw} L.~Bombelli, R.~K.~Koul, J.~Lee and R.~D.~Sorkin,
  ``A Quantum Source of Entropy for Black Holes,''
  Phys.\ Rev.\ D {\bf 34}, 373 (1986);\\
 M.~Srednicki,
  ``Entropy and area,''
  Phys.\ Rev.\ Lett.\  {\bf 71}, 666 (1993)
  [hep-th/9303048].

\bibitem 
 {magic} T.~Jacobson,
  ``Thermodynamics of space-time: The Einstein equation of state,''
  Phys.\ Rev.\ Lett.\  {\bf 75}, 1260 (1995)
  [gr-qc/9504004];\\
T.~Jacobson, ``Gravitation and vacuum entanglement entropy,''
  Int.\ J.\ Mod.\ Phys.\ D {\bf 21}, 1242006 (2012)
  [arXiv:1204.6349 [gr-qc]].

\bibitem 
 {sak} A.~D.~Sakharov,
  ``Vacuum quantum fluctuations in curved space and the theory of gravitation,''
  Sov.\ Phys.\ Dokl.\  {\bf 12}, 1040 (1968)
  [Gen.\ Rel.\ Grav.\  {\bf 32}, 365 (2000)].

\bibitem 
 {rob06} R.~Emparan,
  ``Black hole entropy as entanglement entropy: A Holographic derivation,''
  JHEP {\bf 0606}, 012 (2006)
  [hep-th/0603081].

\bibitem 
 {fur06} D.~V.~Fursaev,
  ``Entanglement entropy in critical phenomena and analogue models of quantum gravity,''
  Phys.\ Rev.\ D {\bf 73}, 124025 (2006)
  [hep-th/0602134].

\bibitem 
 {RS2} L.~Randall and R.~Sundrum,
  ``An Alternative to compactification,''
  Phys.\ Rev.\ Lett.\  {\bf 83}, 4690 (1999)
  [hep-th/9906064].

\bibitem 
 {highRS} R.~Emparan, G.~T.~Horowitz and R.~C.~Myers,
  ``Exact description of black holes on branes,''
  JHEP {\bf 0001}, 007 (2000)
  [hep-th/9911043].

\bibitem 
 {herman} H.~L.~Verlinde,
  ``Holography and compactification,''
  Nucl.\ Phys.\ B {\bf 580}, 264 (2000)
  [hep-th/9906182];\\

\bibitem 
 {gubser} S.~S.~Gubser,
  ``AdS/CFT and gravity,''
  Phys.\ Rev.\ D {\bf 63}, 084017 (2001)
  [hep-th/9912001].

\bibitem 
 {andy} S.~Hawking, J.~M.~Maldacena and A.~Strominger,
  ``de Sitter entropy, quantum entanglement and AdS/CFT,''
  JHEP {\bf 0105}, 001 (2001)
  [hep-th/0002145].

\bibitem 
 {counter} R.~Emparan, C.~V.~Johnson and R.~C.~Myers,
  ``Surface terms as counterterms in the AdS/CFT correspondence,''
  Phys.\ Rev.\ D {\bf 60}, 104001 (1999)
  [hep-th/9903238].

 \bibitem 
 {construct} See, for example:\\
S.~de Haro, S.~N.~Solodukhin and K.~Skenderis, ``Holographic reconstruction of
spacetime and renormalization in the AdS/CFT correspondence,''
  Commun.\ Math.\ Phys.\  {\bf 217}, 595 (2001)
  [arXiv:hep-th/0002230];\\
K.~Skenderis,
  ``Lecture notes on holographic renormalization,''
  Class.\ Quant.\ Grav.\  {\bf 19}, 5849 (2002)
  [arXiv:hep-th/0209067].

\bibitem 
 {aninda} D.~P.~Jatkar and A.~Sinha,
  ``New Massive Gravity and $AdS_4$ counterterms,''
  Phys.\ Rev.\ Lett.\  {\bf 106}, 171601 (2011)
  [arXiv:1101.4746 [hep-th]];\\
K.~Sen, A.~Sinha and N.~V.~Suryanarayana,
  ``Counterterms, critical gravity and holography,''
  Phys.\ Rev.\ D {\bf 85}, 124017 (2012)
  [arXiv:1201.1288 [hep-th]].

\bibitem 
 {string} See, for example:\\
  E.~P.~Verlinde and H.~L.~Verlinde,
  ``RG flow, gravity and the cosmological constant,''
  JHEP {\bf 0005}, 034 (2000)
  [hep-th/9912018];\\
A.~Kehagias,
  ``Exponential and power law hierarchies from supergravity,''
  Phys.\ Lett.\ B {\bf 469}, 123 (1999)
  [hep-th/9906204];\\
A.~Karch and L.~Randall,
  ``Localized gravity in string theory,''
  Phys.\ Rev.\ Lett.\  {\bf 87}, 061601 (2001)
  [hep-th/0105108].
O.~Aharony, O.~DeWolfe, D.~Z.~Freedman and A.~Karch,
  ``Defect conformal field theory and locally localized gravity,''
  JHEP {\bf 0307}, 030 (2003)
  [hep-th/0303249].

\bibitem 
 {induce} T.~Jacobson,
 ``Black hole entropy and induced gravity,''
  gr-qc/9404039;\\
V.~P.~Frolov, D.~V.~Fursaev and A.~I.~Zelnikov,
  ``Statistical origin of black hole entropy in induced gravity,''
  Nucl.\ Phys.\ B {\bf 486}, 339 (1997)
  [hep-th/9607104].

\bibitem 
 {gbee} L.-Y.~Hung, R.~C.~Myers, M.~Smolkin,
  ``On Holographic Entanglement Entropy and Higher Curvature Gravity,''
  JHEP {\bf 1104}, 025 (2011).
  [arXiv:1101.5813 [hep-th]].

\bibitem 
 {casini9} H.~Casini, M.~Huerta and R.~C.~Myers,
  ``Towards a derivation of holographic entanglement entropy,''
  JHEP {\bf 1105}, 036 (2011)
  [arXiv:1102.0440 [hep-th]].

\bibitem 
 {head} M.~Headrick,
  ``Entanglement Renyi entropies in holographic theories,''
  Phys.\ Rev.\  D {\bf 82}, 126010 (2010)
  [arXiv:1006.0047 [hep-th]].

\bibitem 
 {jan} J.~de Boer, M.~Kulaxizi and A.~Parnachev,
  ``Holographic Entanglement Entropy in Lovelock Gravities,''
  JHEP {\bf 1107}, 109 (2011)
  [arXiv:1101.5781 [hep-th]].

\bibitem 
 {feffer} C.~Fefferman and C.~R.~Graham, ``Conformal Invariants,'' in Elie Cartan et les
Math\'ematiques d'aujourd hui (Ast\'erisque, 1985) 95;\\
C.~Fefferman and C.~R.~Graham, ``The Ambient Metric,'' arXiv:0710.0919
[math.DG].

\bibitem 
 {relevant} L.-Y.~Hung, R.~C.~Myers and M.~Smolkin,
  ``Some Calculable Contributions to Holographic Entanglement Entropy,''
  JHEP {\bf 1108}, 039 (2011)
  [arXiv:1105.6055 [hep-th]].

\bibitem 
 {adam} C.~Imbimbo, A.~Schwimmer, S.~Theisen and S.~Yankielowicz,
  ``Diffeomorphisms and holographic anomalies,''
  Class.\ Quant.\ Grav.\  {\bf 17}, 1129 (2000)
  [arXiv:hep-th/9910267];\\
A.~Schwimmer and S.~Theisen,
  ``Entanglement Entropy, Trace Anomalies and Holography,''
  Nucl.\ Phys.\  B {\bf 801}, 1 (2008)
  [arXiv:0802.1017 [hep-th]].

\bibitem 
 {ted0} T.~Jacobson, R.~C.~Myers and ,
  ``Black hole entropy and higher curvature interactions,''
  Phys.\ Rev.\ Lett.\  {\bf 70}, 3684 (1993)
  [hep-th/9305016].

\bibitem 
 {cthem} R.~C.~Myers and A.~Sinha,
  ``Seeing a c-theorem with holography,''
  Phys.\ Rev.\  D {\bf 82}, 046006 (2010)
  [arXiv:1006.1263 [hep-th]];\\
R.~C.~Myers and A.~Sinha,
  ``Holographic c-theorems in arbitrary dimensions,''
  JHEP {\bf 1101}, 125 (2011)
  [arXiv:1011.5819 [hep-th]].

\bibitem 
 {cc} A.~Buchel, J.~Escobedo, R.~C.~Myers, M.~F.~Paulos,
A.~Sinha and M.~Smolkin, ``Holographic GB gravity in arbitrary dimensions,''
  JHEP {\bf 1003} (2010) 111
  [arXiv:0911.4257 [hep-th]].

\bibitem 
 {solo} S.~N.~Solodukhin,
  ``Entanglement entropy, conformal invariance and extrinsic geometry,''
  Phys.\ Lett.\ B {\bf 665}, 305 (2008)
  [arXiv:0802.3117 [hep-th]].

\bibitem 
 {two} H.~Casini and M.~Huerta,
  ``A finite entanglement entropy and the c-theorem,''
  Phys.\ Lett.\  B {\bf 600}, 142 (2004)
  [arXiv:hep-th/0405111];\\
H.~Casini and M.~Huerta,
  ``A c-theorem for the entanglement entropy,''
  J.\ Phys.\ A  {\bf 40}, 7031 (2007)
 [arXiv:cond-mat/0610375].

\bibitem 
 {three} H.~Casini and M.~Huerta,
  ``On the RG running of the entanglement entropy of a circle,''
  Phys.\ Rev.\ D {\bf 85}, 125016 (2012)
  [arXiv:1202.5650 [hep-th]].

\bibitem 
 {maze} H.~Liu and M.~Mezei,
``A Refinement of entanglement entropy and the number of degrees of freedom,''
  arXiv:1202.2070 [hep-th].

\bibitem 
 {flow} R.~C.~Myers and A.~Singh,
  ``Comments on Holographic Entanglement Entropy and RG Flows,''
  JHEP {\bf 1204}, 122 (2012)
  [arXiv:1202.2068 [hep-th]].

\bibitem 
 {zamo} A.~B.~Zamolodchikov,
``Irreversibility of the Flux of the Renormalization Group in a 2D Field
Theory,'' JETP Lett.\  {\bf 43}, 730 (1986)
  [Pisma Zh.\ Eksp.\ Teor.\ Fiz.\  {\bf 43}, 565 (1986)].

\bibitem 
 {Lieb} E.~H.~Lieb and M.~B.~Ruskai,
``Proof of the strong subadditivity of quantum-mechanical entropy,''
  J.\ Math.\ Phys.\  {\bf 14}, 1938 (1973).

\bibitem 
 {Fthem} D.~L.~Jafferis, I.~R.~Klebanov, S.~S.~Pufu and B.~R.~Safdi,
  ``Towards the F-Theorem: N=2 Field Theories on the Three-Sphere,''
  JHEP {\bf 1106}, 102 (2011)
  [arXiv:1103.1181 [hep-th]];\\
I.~R.~Klebanov, S.~S.~Pufu, and B.~R.~Safdi, ``F-Theorem without
Supersymmetry,''
  JHEP {\bf 1110}, 038 (2011)
  [arXiv:1105.4598 [hep-th]].

\bibitem 
 {tadashi} T.~Hirata and T.~Takayanagi,
  ``AdS/CFT and strong subadditivity of entanglement entropy,''
  JHEP {\bf 0702}, 042 (2007)
  [hep-th/0608213].

\bibitem 
 {important} C.~G.~Callan, Jr. and F.~Wilczek,
  ``On geometric entropy,''
  Phys.\ Lett.\ B {\bf 333}, 55 (1994)
  [hep-th/9401072];\\
L.~Susskind and J.~Uglum, ``Black hole entropy in canonical quantum gravity and
superstring theory,''
  Phys.\ Rev.\ D {\bf 50}, 2700 (1994)
  [hep-th/9401070].

\bibitem 
 {beke0} J.~D.~Bekenstein, ``A Universal Upper Bound on the Entropy to Energy Ratio for
Bounded Systems,''  Phys.\ Rev.\ D {\bf 23}, 287 (1981).

\bibitem 
 {bousso} R.~Bousso, ``A Covariant entropy conjecture,''
  JHEP {\bf 9907}, 004 (1999)
  [hep-th/9905177];\\
R.~Bousso, ``The Holographic principle,''
  Rev.\ Mod.\ Phys.\  {\bf 74}, 825 (2002)
  [hep-th/0203101].

\bibitem 
 {flan} E.~E.~Flanagan, D.~Marolf and R.~M.~Wald,
``Proof of classical versions of the Bousso entropy bound and of the
generalized second law,''
  Phys.\ Rev.\ D {\bf 62}, 084035 (2000)
  [hep-th/9908070].

\bibitem 
 {beke1} H.~Casini, ``Relative entropy and the Bekenstein bound,''  Class.\ Quant.\
Grav.\  {\bf 25}, 205021 (2008)  [arXiv:0804.2182 [hep-th]];\\
D.~D.~Blanco, H.~Casini, L.-Y.~Hung and R.~C.~Myers, in preparation.

\bibitem 
 {furx} D.~V.~Fursaev,
  ``Entanglement entropy in quantum gravity and the Plateau groblem,''
  Phys.\ Rev.\ D {\bf 77}, 124002 (2008)
  [arXiv:0711.1221 [hep-th]].

\bibitem 
 {furx2} D.~V.~Fursaev,
  ```Thermodynamics' of Minimal Surfaces and Entropic Origin of Gravity,''
  Phys.\ Rev.\ D {\bf 82}, 064013 (2010)
  [Erratum-ibid.\ D {\bf 86}, 049903 (2012)]
  [arXiv:1006.2623 [hep-th]].

\bibitem 
 {Carea} H.~Casini,
  ``Geometric entropy, area, and strong subadditivity,''
  Class.\ Quant.\ Grav.\  {\bf 21}, 2351 (2004)
  [hep-th/0312238].

\bibitem 
 {lovel} D.~Lovelock, ``The Einstein tensor and its generalizations,''
  J.\ Math.\ Phys.\  {\bf 12}, 498 (1971);\\
D.~Lovelock, ``Divergence-free tensorial concomitants,'' Aequationes Math. {\bf
4}, 127 (1970).

\bibitem 
 {highc} S.~Nojiri and S.D.~Odintsov,
``On the conformal anomaly from higher derivative gravity in AdS/CFT
correspondence,''
  Int.\ J.\ Mod.\ Phys.\  A {\bf 15}, 413 (2000)
  [arXiv:hep-th/9903033];\\
M.~Blau, K.S.~Narain and E.~Gava, ``On subleading contributions to the AdS/CFT
trace anomaly,''
  JHEP {\bf 9909}, 018 (1999)
  [arXiv:hep-th/9904179].

\bibitem 
 {EtasGB} M.~Brigante, H.~Liu, R.~C.~Myers, S.~Shenker and S.~Yaida,
  ``Viscosity Bound Violation in Higher Derivative Gravity,''
  Phys.\ Rev.\  D {\bf 77}, 126006 (2008)
  [arXiv:0712.0805 [hep-th]];\\
A.~Buchel and R.~C.~Myers,
  ``Causality of Holographic Hydrodynamics,''
  JHEP {\bf 0908}, 016 (2009)
  [arXiv:0906.2922 [hep-th]];\\
  J.~de Boer, M.~Kulaxizi and A.~Parnachev,
  ``AdS$_7$/CFT$_6$, Gauss-Bonnet Gravity, and Viscosity Bound,''
  JHEP {\bf 1003}, 087 (2010)
  [arXiv:0910.5347 [hep-th]];\\
X.~O.~Camanho and J.~D.~Edelstein,
  ``Causality constraints in AdS/CFT from conformal collider physics and
  Gauss-Bonnet gravity,''
  JHEP {\bf 1004}, 007 (2010)
  [arXiv:0911.3160 [hep-th]];\\
 J.~de Boer, M.~Kulaxizi and A.~Parnachev,
  ``Holographic Lovelock Gravities and Black Holes,''
  JHEP {\bf 1006}, 008 (2010)
  [arXiv:0912.1877 [hep-th]];\\
D.~M.~Hofman,
  ``Higher Derivative Gravity, Causality and Positivity of Energy in a UV
  complete QFT,''
  Nucl.\ Phys.\  B {\bf 823}, 174 (2009)
  [arXiv:0907.1625 [hep-th]];\\
X.~O.~Camanho and J.~D.~Edelstein,
  ``Causality in AdS/CFT and Lovelock theory,''
  JHEP {\bf 1006}, 099 (2010)
  [arXiv:0912.1944 [hep-th]].

\bibitem 
 {surf} R.~C.~Myers,
  ``Higher Derivative Gravity, Surface Terms And String Theory,''
  Phys.\ Rev.\ D {\bf 36}, 392 (1987);\\
R.~Olea, ``Mass, angular momentum and thermodynamics in four-dimensional
Kerr-AdS black holes,''
  JHEP {\bf 0506}, 023 (2005)
  [hep-th/0504233].

\bibitem 
 {GBghost} D.~G.~Boulware and S.~Deser,
  ``String Generated Gravity Models,''
  Phys.\ Rev.\ Lett.\  {\bf 55} (1985) 2656.

\bibitem 
 {old1} R.~C.~Myers and B.~Robinson,
  ``Black Holes in Quasi-topological Gravity,''
  JHEP {\bf 1008}, 067 (2010)
  [arXiv:1003.5357 [gr-qc]];\\
R.~C.~Myers, M.~F.~Paulos and A.~Sinha,
  ``Holographic studies of quasi-topological gravity,''
  JHEP {\bf 1008}, 035 (2010)
  [arXiv:1004.2055 [hep-th]].

\bibitem 
 {yale} A.~Yale, ``Simple counterterms for asymptotically AdS
 spacetimes in Lovelock gravity,''
  Phys.\ Rev.\ D {\bf 84}, 104036 (2011)
  [arXiv:1107.1250 [gr-qc]].

\end{thebibliography}
\end{document}